\documentclass[
twocolumn,
english,
aps,
pra,
longbibliography,
superscriptaddress,
amsmath,
amssymb,
floatfix,
]{revtex4-2}
\usepackage{graphicx}% Include figure files
\usepackage{dcolumn}% Align table columns on decimal point
\usepackage{bm}% bold math
\usepackage{physics}
\usepackage{hyperref}
\usepackage{subcaption}
\usepackage{quantikz}
\usepackage{tikz}
\definecolor{riverlane_green}{RGB}{0, 150, 143}

\usepackage{soul}

\begin{document}

\preprint{APS/123-QED}

\title{\textbf{Quantum Data Learning of Topological-to-Ferromagnetic Phase Transitions in the 2+1D Toric Code Loop Gas Model} 
}% 

\author{Shamminuj Aktar}
\email{saktar@lanl.gov} % corresponding author
\affiliation{Computing and Artificial Intelligence Division (CAI-3), Los Alamos National Laboratory, Los Alamos, New Mexico 87545, USA}

\author{Rishabh Bhardwaj}
\affiliation{Computing and Artificial Intelligence Division (CAI-3), Los Alamos National Laboratory, Los Alamos, New Mexico 87545, USA}

\author{Andreas B\"artschi}
\affiliation{Computing and Artificial Intelligence Division (CAI-3), Los Alamos National Laboratory, Los Alamos, New Mexico 87545, USA}

\author{Tanmoy Bhattacharya}
\affiliation{Theoretical Division (T-2), Los Alamos National Laboratory, Los Alamos, New Mexico 87545, USA}

\author{Stephan Eidenbenz}
\affiliation{Computing and Artificial Intelligence Division (CAI-3), Los Alamos National Laboratory, Los Alamos, New Mexico 87545, USA}

\date{\today}% It is always \today, today,

\begin{abstract}
    Quantum data learning (QDL) provides a framework for extracting physical insights directly from quantum states, bypassing the need for any identification of the classical observable of the theory. A central challenge in many-body physics is that the identity of quantum phases, especially those with topological order, are often inaccessible through local observables or simple symmetry-breaking diagnostics. Here, we apply QDL techniques to the \(2{+}1\)-dimensional toric-code loop-gas model in a magnetic field. Ground states are generated across multiple lattice sizes using a parametrized loop-gas circuit (PLGC) with a variational quantum-eigensolver (VQE) approach. We then train a quantum convolutional neural network (QCNN) across the full field-parameter range to perform phase classification and capture the overall phase structure. We also employ a physics-aware training protocol that excludes the near-critical region (\(0.2 \le x \le 0.4\)) around \(x_c \approx 0.25\), the phase-transition point estimated by quantum Monte Carlo, reserving this window for testing to evaluate the ability of the model to learn the phase transition. In parallel, we implement an unsupervised quantum \(k\)-means method based on state overlaps, which partitions the dataset into two phases without prior labeling. 
    Our supervised QDL approach recovers the phase structure and accurately locates the phase transition, in close agreement with previously reported values; the unsupervised QDL approach recovers the phase structure and locates the phase transition with a small offset as expected in finite volumes; both QDL methods outperform classical alternatives.
    %\se{original sentence that I rephrased, but still just leave here: Both supervised and unsupervised QDL approaches successfully recover the phase structure and accurately locate the phase transition, in close agreement with previously reported values.} 
    These findings establish QDL as an effective framework for characterizing topological quantum matter, studying finite volume effects, and probing phase diagrams of higher-dimensional systems.
\end{abstract}

%\keywords{Quantum Data Learning, Quantum Convolutional Neural Network, Parametrized Loop-Gas Circuit, Toric Code, Topological Order, Quantum Phase Transition}%Use showkeys class option if keyword
\maketitle

%\tableofcontents

\section{Introduction}
\label{sec:intro}

The study of quantum many-body systems lies at the heart of condensed-matter and high-energy physics~\cite{Meglio2024quantum}. Strong correlations among many particles lead to emergent behavior that cannot be reduced to single-particle descriptions, underpinning phenomena such as high-temperature superconductivity, quantum spin liquids, and topological order. Classical computational approaches, including density functional theory (DFT)~\cite{hohenberg1964,kohn1999nobel}, quantum Monte Carlo (QMC)~\cite{Ceperley1986quantum,becca2017quantum}, and the density matrix renormalization group (DMRG)~\cite{PhysRevLett.69.2863,PhysRevB.48.10345}, have provided profound insights into correlated quantum systems. However, each of these methods face intrinsic limitations: DFT relies on approximate exchange–correlation functionals that break down in strongly correlated regimes; QMC is constrained by the fermion sign problem and unstable analytic continuation; and tensor-network approaches are efficient only for low-entanglement states, with poor scaling in higher dimensions or real-time dynamics. More recently, classical machine learning has been leveraged for phase classification and order-parameter discovery in many-body systems~\cite{RevModPhys.91.045002,Carrasquilla2020,Deng2017machine,Carleo2017solving,Wang2016discovering,RodriguezNieva2019,Glasser1018neural}. Yet the exponential growth of Hilbert space, extensive post-processing required, and absence of local order parameters remain major obstacles to studying large many-body systems. These limitations motivate the exploration of alternative strategies that exploit quantum computational resources. As argued by Huang \textit{et al.}~\cite{huang2022provably}, quantum models can, in principle, efficiently extract physical information inaccessible to classical learners, suggesting a new path for studying correlated matter.

Quantum machine learning (QML), a rapidly developing framework, combines the expressive power of parameterized quantum circuits with the adaptive strategies of modern learning algorithms~\cite{biamonte2017,schuld2019,mangini2021quantum, cerezo2022challenges, cerezo2021variational}. Most QML applications to date have focused on encoding classical data into quantum states, which are subsequently processed by hybrid quantum–classical models for classification or regression tasks~\cite{havlivcek2019,schuld2021,benedetti2019}. This paradigm has enabled progress in diverse domains, including quantum chemistry~\cite{von2020exploring}, materials science~\cite{rupp2012}, and high-energy physics~\cite{blance2021,guan2021quantum,wu2022challenges}. However, it inherits the cost of designing efficient encodings and does not fully utilize the inherent quantum structure of many-body states. Recently, a new direction, \emph{quantum data learning} (QDL), has been proposed, in which quantum models learn directly from quantum-mechanical data. This eliminates the need for classical encodings and enables direct manipulation of quantum wavefunctions.

QDL techniques are naturally suited for studying high-energy physics. Recent developments in collider experiments are enabling the direct probing of quantum states through measurable observables such as angular distributions, momentum correlations, and precision timing, offering access to entanglement entropy and quantum coherence~\cite{Fabbrichesi:2021npl,Severi:2021cnj,Aguilar-Saavedra:2022uye,Hentschinski:2024gaa}. Moreover, quantum tomography has already been demonstrated in systems of top quarks, photon pairs, and entangled mesons~\cite{Afik:2020onf,Aoude:2022imd,Fabbrichesi:2022ovb,Ehataht:2023zzt}, together with quantum-information analyses in quantum chromodynamics~\cite{Hentschinski:2024gaa}. With upcoming detector upgrades at the Large Hadron Collider (LHC) expected to provide real-time access to such quantum observables, QDL offers a promising framework for extracting physical insights directly from high-dimensional quantum data.

Earlier QDL studies have focused on abstract or low-dimensional systems. Nagano \textit{et al.}~\cite{nagano2023quantum} employed quantum convolutional neural network (QCNN) architectures on \((1{+}1)\)-dimensional systems, including the Schwinger model and the \(\mathbb{Z}_2\) lattice gauge theory, identifying quantum phases and extracting coupling parameters in parton-shower-inspired simulations. These studies demonstrated that QDL can uncover nontrivial structures without explicit order parameters, motivating extensions to higher-dimensional systems.

\begin{figure*}[t!]
    \includegraphics[width=1\textwidth]{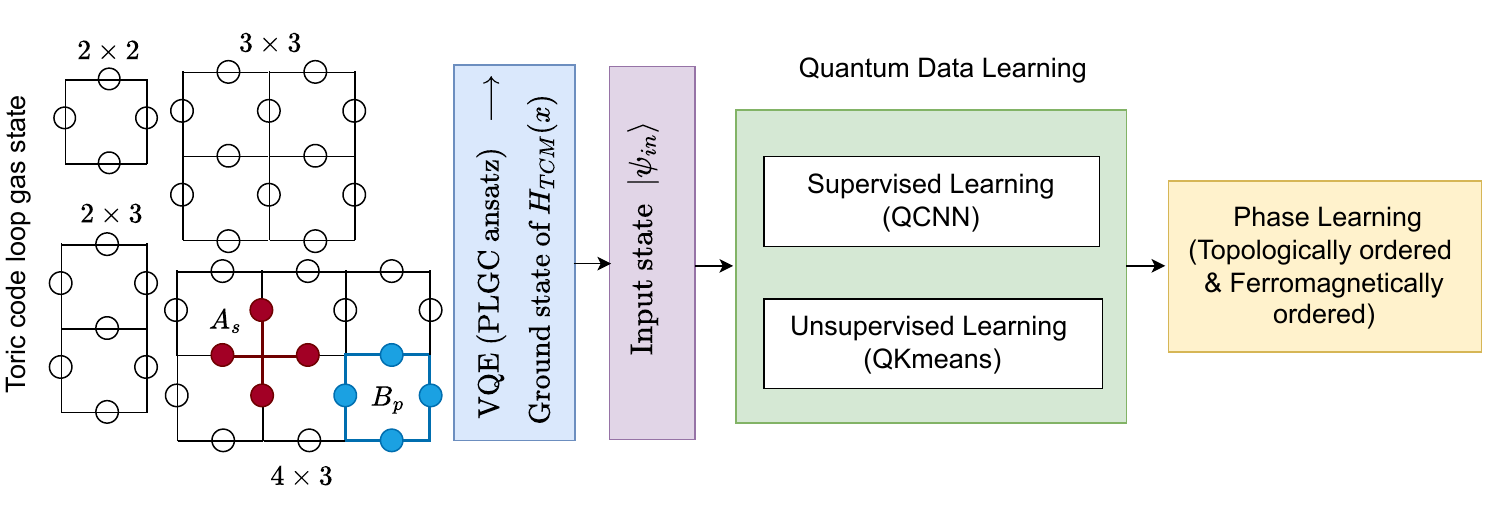}
    \caption{Overview of the workflow for quantum data learning of phases in the \(2{+}1\)-dimensional toric-code loop-gas model under a magnetic field. Ground states of \(H{(x)}\) are approximated by a variational quantum eigensolver (VQE) using the parametrized loop-gas circuit (PLGC) on finite lattices. The resulting states \(|\psi_x\rangle\) are processed using QDL techniques to learn the phase structure and distinguish between topologically ordered and ferromagnetically ordered phases.}
    \label{fig:toric_code_qdl}
\end{figure*}

In this work, we investigate QDL techniques in the \(2{+}1\)-dimensional toric-code model in a magnetic field, a paradigmatic system that hosts a transition between a topological phase and a ferromagnetic phase~\cite{trebst2007,sun2023parametrized}. The toric code model captures the essential physics of confinement while remaining accessible to both analytical control and efficient simulation. It exhibits topological order, emergent anyonic excitations, and a confinement–deconfinement transition, making it a minimal yet rich platform for probing the interplay between quantum correlations and phase structure. From a practical standpoint, the model is simple enough to be tractable on classical computers and implementable on near-term quantum devices, while also serving as a meaningful stepping stone toward more complex gauge theories, including non-Abelian models and ultimately quantum chromodynamics. Its relevance extends beyond theory: recent advances in programmable quantum simulators, such as Rydberg-atom arrays, have realized quantum dimer models with nonlocal string operators, directly observing hallmarks of toric-code physics including anyonic excitations, coherent topological sectors, and loop-parity measurements~\cite{Semeghini_2021,Weimer_2010}.

To probe the phase structure of the toric-code model, we generate ground states using a parametrized loop-gas circuit (PLGC) within a variational quantum eigensolver (VQE), following the approach of Sun \textit{et al.}~\cite{sun2023parametrized}, for lattice sizes up to \(4{\times}3\). First, we train QCNNs across the full range of the field parameter and observe nearly perfect classification accuracy, demonstrating that phase information can be extracted directly from the quantum ground states. We then adopt a physics-aware strategy in which states far from the QMC-estimated transition at \(x_c \approx 0.25\)~\cite{trebst2007, Wu2012phase} are used for training, while those in the region (\(0.2 \le x \le 0.4\)) are reserved for testing. From the QCNN decision function, we extract phase transition as flip interval estimators \(x_c'\) for each lattice size and observe systematic convergence toward \(x_c\) as system size increases. In parallel, we implement an unsupervised quantum \(k\)-means algorithm based on fidelity overlaps, which partitions the data into two phases without labels. The resulting cluster flip interval estimators \(\hat{x}_c\) capture the overall phase boundary and broadly follow the QCNN-predicted transition points \(x_c'\), though with a small offset. 

Finally, we benchmark QDL in learning phase transition of the toric-code model against classical machine-learning approaches, including convolutional neural networks (CNNs) and logistic regression. Both the CNN and logistic regression models were trained separately on the VQE ansatz parameters and on the quantum state amplitudes as input features. While classical models capture the broad phase structure, QCNNs consistently yield more reliable transition points, particularly for the \(4{\times}3\) lattice where the estimate is approaching the QMC estimate. Moreover, when varying the training-set size, QCNNs are more sample-efficient than other baselines in learning the phase transition. Finite-size scaling further shows that the QCNN-extrapolated transition \(x_c(\infty) = 0.2518 \pm 0.026\) converges closely to the QMC benchmark \(x_c \approx 0.25\).

The rest of this work is organized as follows. Section~\ref{sec:method} outlines the ground-state generation using the parametrized loop-gas ansatz 
and describes the quantum data learning framework, including supervised QCNN and unsupervised quantum \(k\)-means models. 
Section~\ref{sec:results} presents the experimental results on learning phase transition, comparing QDL-based phase learning with classical machine-learning methods and performing finite-size scaling. Finally, Section~\ref{sec:conclusion} summarizes the findings and discusses future directions. 
%\se{Add a sentence on the appendices?)}

\section{Methodology}
\label{sec:method}
Fig.~\ref{fig:toric_code_qdl} provides an overview of the QDL framework applied to the \(2{+}1\)-dimensional toric-code loop-gas model under a magnetic field. 
First, ground states are prepared within a variational quantum eigensolver (VQE) setup across different lattice geometries. The resulting states serve as quantum inputs \(|\psi_x\rangle\) for the learning stage. These input states are then processed either by QCNNs for supervised phase-transition learning or by quantum \(k\)-means models for unsupervised clustering. Both approaches enable the identification of the transition between the topologically ordered and ferromagnetically ordered phases. 

\subsection{Toric Code Loop-Gas Model}
\label{sec:toric-code-model}

Topological phases represent a remarkable class of quantum matter whose low-energy properties remain robust against local perturbations and cannot be characterized by conventional symmetry-breaking order parameters. These phases have attracted substantial interest in condensed-matter and quantum-information science due to their inherent stability, which makes them promising platforms for fault-tolerant quantum computation~\cite{trebst2007}. A paradigmatic example of a topologically ordered system is Kitaev's toric code, an exactly solvable lattice model described by the Hamiltonian~\cite{kitaev2010topological}
\begin{equation}
    H_{\mathrm{TCM}} = -\sum_{s} A_s - \sum_{p} B_p ,
\end{equation}
where $A_s = \prod_{i \in s}\sigma_i^x$ (star operators acting on edges adjacent to vertex $s$) and $B_p = \prod_{i \in p}\sigma_i^z$ (plaquette operators acting on edges surrounding plaquette $p$) mutually commute. Fig.~\ref{fig:toric_code_qdl} illustrates these operators on a square lattice, highlighting a star operator $A_s$ \textit{(red)} and a plaquette operator $B_p$ \textit{(blue)}. The ground state of the toric code is a coherent superposition of closed-loop spin configurations defined on the lattice edges, commonly referred to as a quantum loop gas.

\begin{figure*}[t!]
    \centering
    \includegraphics[width=0.45\textwidth]{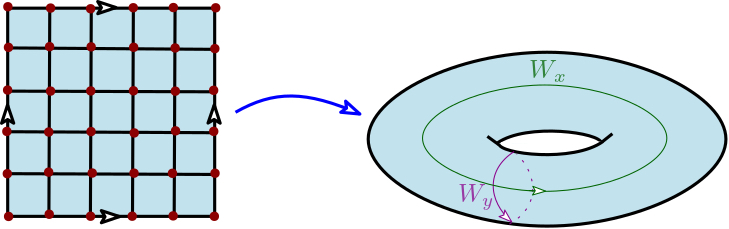}
    \hfill
    \includegraphics[width=0.45\textwidth]{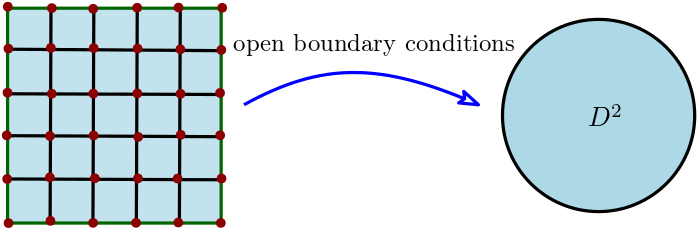}
    \caption{Toric geometry and boundary effects in the toric code model (TCM). 
    (\textit{left}) Construction of the toric geometry under periodic boundary conditions, yielding a manifold with two non-contractible Wilson loops $W_x$ and $W_y$ that define distinct topological sectors of the ground state and encode the intrinsic $\mathbb{Z}_2$ topological order. 
    (\textit{right}) Under open boundary conditions, the finite square lattice is topologically equivalent to a disk \(D^2\), eliminating non-contractible loops and hence the global ground-state degeneracy while preserving the bulk $\mathbb{Z}_2$ order.}
    \label{fig:tcm_geometry}
\end{figure*}

Under periodic boundary conditions (PBC), opposite edges of the lattice are identified, mapping the system onto a torus \(T^2\). This topology eliminates boundary effects and associated edge excitations (anyons), leaving only intrinsic topological properties. A key consequence is a degenerate ground-state manifold composed of distinct topological sectors. These sectors correspond to elements of the first homology group \(H_1(T^2,\mathbb{Z}_2)\) and are characterized by the \(\mathbb{Z}_2\) winding parities along noncontractible loops in the \(x\) and \(y\) directions. Specifically, the winding sectors are defined by Wilson-loop operators
\begin{equation*}
    W^a_{x/y} = \prod_{i \in c_{x/y}} \sigma_i^a ,
\end{equation*}
where \(c_{x/y}\) denotes non-contractible cuts along the \(x/y\) directions, and the index \(a = x,z\) labels the electric and magnetic sectors, respectively. This construction is illustrated schematically in Fig.~\ref{fig:tcm_geometry} \textit{(left)}. The ground states obtained by coherent superpositions of these sectors are immune to mixing by any local perturbations.

When open boundary conditions (OBC) are applied, the lattice effectively reduces to a disk shown in Fig.~\ref{fig:tcm_geometry} \textit{(right)}. As a result, the topology-protected ground-state degeneracy is lost. The bulk topological order, however, persists: loop configurations behave as in the toric case, spins remain long-range entangled, and the system supports deconfined anyons whose boundaries can absorb or condense them. The phase retains a nonzero topological entanglement entropy \(\gamma = \ln 2\).\footnotetext{This quantity is a topological invariant and is independent of lattice geometry.} Depending on the boundary termination (``rough'' or ``smooth''), edge anyon condensation lifts the global degeneracy, while in geometries with multiple boundaries (e.g., a cylinder or punctured plane) a reduced, boundary-dependent degeneracy reemerges. In essence, OBC on finite systems remove the global degeneracy and introduce anyonic edge modes but do not destroy the underlying \(\mathbb{Z}_2\) topological order. In the present work, we employ OBC, which allow for explicit treatment of edge effects and simplify the implementation on finite lattices by removing nonlocal stabilizer constraints and periodic wrap-around couplings required under PBC. 

%Under OBC, the number of independent plaquette and star operators equals the number of spins, yielding a uniquely defined ground state without enforcing global loop constraints, which greatly reduces the computational overhead~\cite{kitaev2010topological,trebst2007}. 
%\tb{Cite literature or otherwise explain why OBC is simpler to implement?}

External perturbations such as a uniform magnetic field introduce an effective loop tension that destabilizes the topological phase. When the field is sufficiently strong, it selects a single, topologically trivial sector and drives a continuous second-order quantum phase transition into a ferromagnetically ordered state~\cite{trebst2007,sun2023parametrized}. Physically, this transition corresponds to the condensation of magnetic vortices and the confinement of emergent \(\mathbb{Z}_2\) charges, providing a canonical example of the breakdown of topological order.

Following the methodology introduced by Sun \textit{et al.}~\cite{sun2023parametrized}, we study the toric-code model in a uniform longitudinal magnetic field. Qubits are placed on the lattice edges under OBC, and the Hamiltonian interpolates continuously between the topologically ordered and ferromagnetic limits,
\begin{equation}
    H(x) = (1 - x)H_{\mathrm{TCM}} - x \sum_{i=1}^{N} \sigma_i^z ,
\end{equation}
where the parameter \(x \in [0,1]\) tunes the relative strength between the stabilizer terms \((A_s, B_p)\), which favor topological order at \(x=0\), and the magnetic field term \((\sigma_i^z)\), which favors ferromagnetic alignment at \(x=1\). Quantum Monte Carlo simulations estimate the critical transition point between these phases at \(x_c \approx 0.25\) in the thermodynamic limit~\cite{trebst2007}. 
%\tb{I don't think \cite{Wu2012phase} were the first to get this, they seem to cite previous literature on this point.  It is also important to check whether the 0.25 (or 0.333 in their parameterization) is actually an analytic answer, or only numerical.}
%This quantum phase transition is entirely governed by the ground state. 
Traditional QMC analyses identify the critical point using a local order parameter—typically the average magnetization along the \(z\)-axis,
\begin{equation}
    \langle m_z\rangle = \frac{1}{N} \left\langle \sum_{i=1}^{N} \sigma_i^z \right\rangle ,
    \label{eq:magnetization_per_qubit}
\end{equation}
and by examining the Binder cumulant
\begin{equation*}
    Q(m_z) = \frac{\langle m_z^2\rangle^2}{\langle m_z^4\rangle} ,
\end{equation*}
together with its finite-size scaling behavior near \(x_c\). In contrast, the present work aims to predict this critical point by learning directly from the ground states prepared on a quantum processor. % \st{directly from the ground state itself}}. %without the need oinvoking any explicit local order parameter.

\subsection{Toric-Code Ground-State Preparation}
\label{sec:toric-code-ground}

To study quantum phases and their transitions, it is essential to obtain accurate approximations of the ground states of the system Hamiltonian. In principle, such ground states can be prepared using quantum phase estimation (QPE)~\cite{nielsen2000quantum} or computed via exact diagonalization (ED); however, both approaches are resource intensive and become impractical on hardware available today even for modest lattice sizes. In this work, we instead employ the VQE, which offers a hardware-efficient framework for approximating ground states.

We adopt the physics-motivated parametrized loop-gas circuit (PLGC) introduced in Ref.~\cite{sun2023parametrized}. The construction begins from the trivial product state \(\ket{0}^{\otimes N}\) and applies one unitary per plaquette, replacing the Hadamard gate in the exact toric-code preparation with a tunable rotation \(R_y(\theta_p)\). The resulting variational state takes the form
\begin{equation}
    \ket{\psi_x(\boldsymbol{\theta})}
    = \prod_{p=1}^{N_p}
    \left[
    \cos\!\left(\tfrac{\theta_p}{2}\right) I
    + \sin\!\left(\tfrac{\theta_p}{2}\right) B_p
    \right]
    \ket{0}^{\otimes N},
\end{equation}
where \(B_p = \prod_{i \in p} \sigma_i^x\) is the plaquette operator acting on the four edges surrounding plaquette \(p\). This circuit generates weighted superpositions of closed-loop configurations, interpolating between the equal-weight loop gas at \(\theta_p = \pi/2\) and the trivial product states at \(\theta_p = 0\) or \(\pi\). The PLGC thus confines the variational search space within the physically relevant loop-gas subspace while retaining tunability across different field regimes~\cite{sun2023parametrized}.

For each lattice size, optimized ground states are prepared across a grid of field values \(x\). The VQE cost function is the energy expectation value
\begin{equation}
    E(\boldsymbol{\theta};x)
    = \bra{\psi(\boldsymbol{\theta})}
    H_{\mathrm{TCM}}(x)
    \ket{\psi(\boldsymbol{\theta})},
\end{equation}
which is estimated from quantum measurements on the PLGC ansatz. Details of the VQE training procedure are provided in Appendix~\ref{appx:vqe}. After generating the ground states, we assign phase labels to the states based on our aproximate knowledge of the phase transition point: states with \(x \le x_c\) are identified with the topologically ordered phase (\(-1\)), while those with \(x > x_c\) correspond to the ferromagnetically ordered phase (\(+1\)), where \(x_c \approx 0.25\) is the QMC-estimated critical point in the thermodynamic limit~\cite{trebst2007, Wu2012phase}. In Subsection~\ref{sec:qcnn}, we describe a quantum data learning method to refine this input approximate \(x_c\) to a precise finite-volume transition point, and in Subsection~\ref{sec:qkmeans}, we devise an alternate method to find the transition point without using any such input.

\subsection{Quantum Data Learning Framework}
\label{sec:qdl-framework}
\subsubsection{Supervised Learning: QCNN}
\label{sec:qcnn}

For supervised phase classification, we employ quantum convolutional neural networks (QCNNs), a hierarchical circuit ansatz inspired by multiscale entanglement renormalization. QCNNs possess a logarithmic-depth architecture that is provably resilient to barren plateaus, enabling efficient training on near-term quantum devices~\cite{pesah2021absence}. In addition, the number of variational parameters grows only polylogarithmically with system size, and the architecture is naturally compatible with fault-tolerant implementations. Previous studies have demonstrated that QCNNs are powerful tools for quantum phase recognition: Cong \textit{et al.\hbox{}} first proposed QCNNs for identifying topological phases in spin chains~\cite{cong2019}, and more recently, Nagano \textit{et al.\hbox{}} applied them to lattice gauge theories and high-energy physics simulations~\cite{nagano2023quantum}, successfully identifying confinement, deconfinement, and coupling parameters directly from quantum states. These successes motivate their application to the toric-code model considered here. 
%\tb{\LaTeX\ note: period used for abbreviation except after a capital letter needs something special to fix spacing. See what I have done.}

Following the construction in Ref.~\cite{nagano2023quantum}, our QCNN alternates convolution and pooling layers in a recursive manner, reducing the number of qubits at each stage while retaining global information about the system (see Appendix~\ref{appx:qcnn} for the QCNN architecture diagram and the circuit implementations of each convolution and pooling block). The ground states $\ket{\psi_x}$ prepared by the VQE serve directly as inputs to the QCNN. The first convolution layer consists of two-qubit \(SU(4)\) unitaries arranged in a brickwork pattern across the full register, efficiently entangling local degrees of freedom. The subsequent pooling layers employ parameterized single-qubit rotations followed by CNOT gates to project out qubits, effectively renormalizing the system while preserving some nonlocal correlations. The following convolution layers act on fewer qubits by pairing only adjacent ones, and this coarse-graining process continues until only two qubits remain before the final pooling step. The last pooling block outputs a single qubit, which is measured in the \(Z\) basis to yield the QCNN output. 

The resulting measurement outcome serves as a phase classifier, distinguishing the \(\mathbb{Z}_2\) topologically ordered phase from the ferromagnetically ordered phase. 
Formally, the QCNN prediction for an input quantum state \(|\psi_x\rangle\) is defined as the expectation value of the Pauli-\(Z\) operator on the final output qubit after processing through the QCNN unitary circuit \(U_{\mathrm{QCNN}}(\boldsymbol{\phi})\):
\begin{equation}
    y_{\mathrm{out}}(\boldsymbol{\phi})
    = \mathrm{Tr}\!\left[
        Z_{\mathrm{out}} \,
        U_{\mathrm{QCNN}}(\boldsymbol{\phi})
        \, |\psi_x\rangle\!\langle\psi_x| \,
        U_{\mathrm{QCNN}}^\dagger(\boldsymbol{\phi})
      \right],
    \label{eq:qcnn_output}
\end{equation}
where \(Z_{\mathrm{out}}\) acts on the remaining qubit, and \(\boldsymbol{\phi}\) denotes the set of trainable parameters in the QCNN. 
The output \(y_{\mathrm{out}} \in [-1, 1]\) serves as a continuous phase indicator: \(y_{\mathrm{out}} \leq 0\) corresponds to the topologically ordered phase $(-1) $, while \(y_{\mathrm{out}} > 0\) corresponds to the ferromagnetically ordered phase $(+1)$.

To identify the phase boundary, we analyze the QCNN decision function as a function of the tuning parameter \(x\). 
We employ a robust estimator of the transition point by defining \(x_c'\) as the central value of the interval \([x_{j_{\min}}, x_{j_{\max}}]\), where \(j_{\min}\) is the largest index such that all samples with \(x_i \leq x_{j_{\min}}\) share the same QCNN predicted phase label, and \(j_{\max}\) is the smallest index such that all samples with \(x_i \geq x_{j_{\max}}\) share the opposite phase label. 
The estimated flip interval point is then
\begin{equation}
    x_c' = \tfrac{1}{2}\big(x_{j_{\min}} + x_{j_{\max}}\big),
\end{equation}
with an uncertainty of \(\Delta x_c' = (x_{j_{\max}} - x_{j_{\min}})/2\) associated to it. 

%\tb{The word variance is misused here.  I would say ``with an uncertainty of \(\Delta x_c' = (x_{j_{\max}} - x_{j_{\min}})/2\) associated to it.''  I am OK with replacing the \(2\) by \(2\sqrt3\) if desired.}
%This procedure provides a stable and noise-resilient estimate of the finite-size phase transition point. \tb{I would delete this last statement.  No point emphasizeing this: we have described what we did, called it a stable estimator. Be done with it.}%within the supervised QDL framework.%, ensuring consistency. % even when the QCNN outputs exhibit nonmonotonic behavior or statistical fluctuations.

 \begin{figure*}[t!]
    \centering
    \begin{subfigure}[t]{0.495\linewidth}
        \centering
        \includegraphics[width=\linewidth]{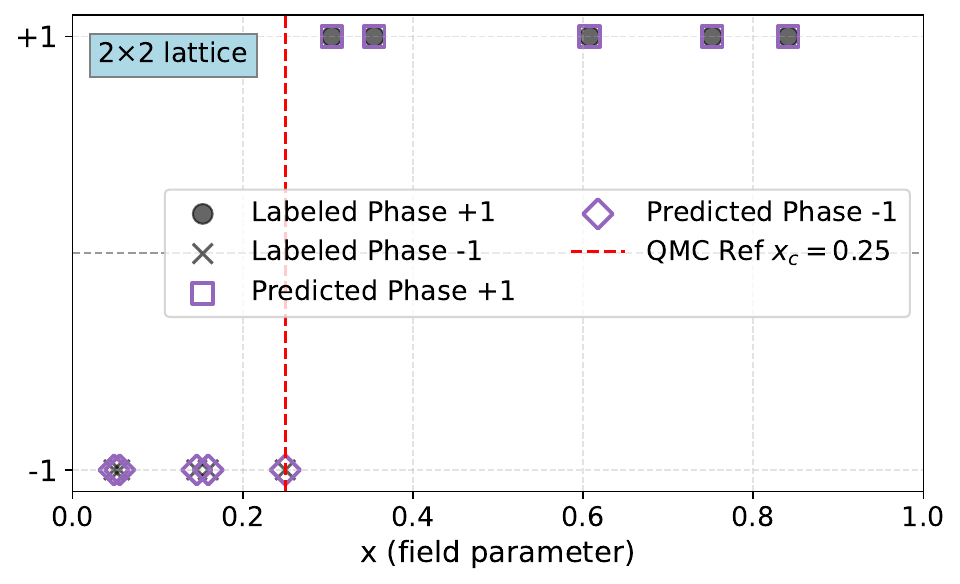}
    \end{subfigure}\hfill
    \begin{subfigure}[t]{0.495\linewidth}
        \centering
        \includegraphics[width=\linewidth]{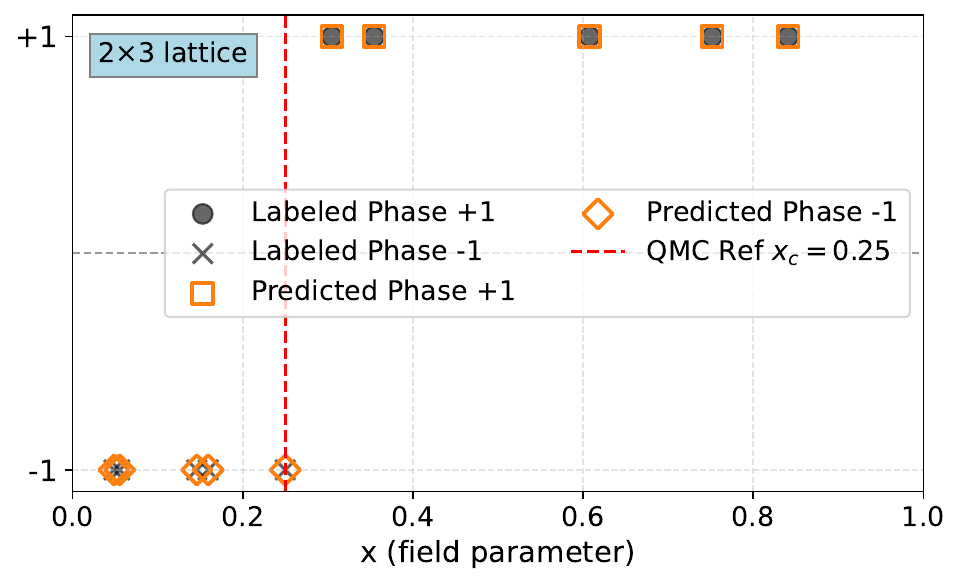}
    \end{subfigure}

    \begin{subfigure}[t]{0.495\linewidth}
        \centering
        \includegraphics[width=\linewidth]{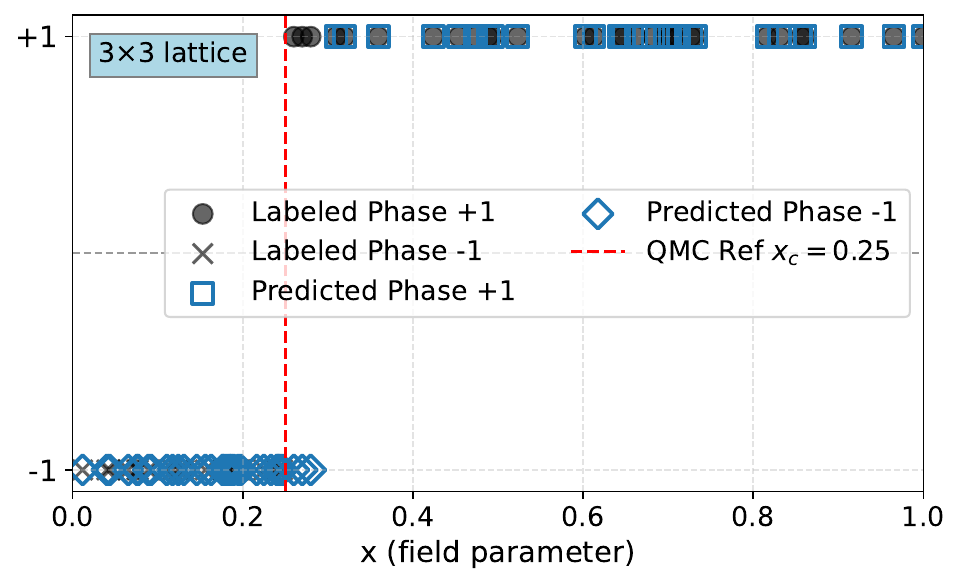}
    \end{subfigure}\hfill
    \begin{subfigure}[t]{0.495\linewidth}
        \centering
        \includegraphics[width=\linewidth]{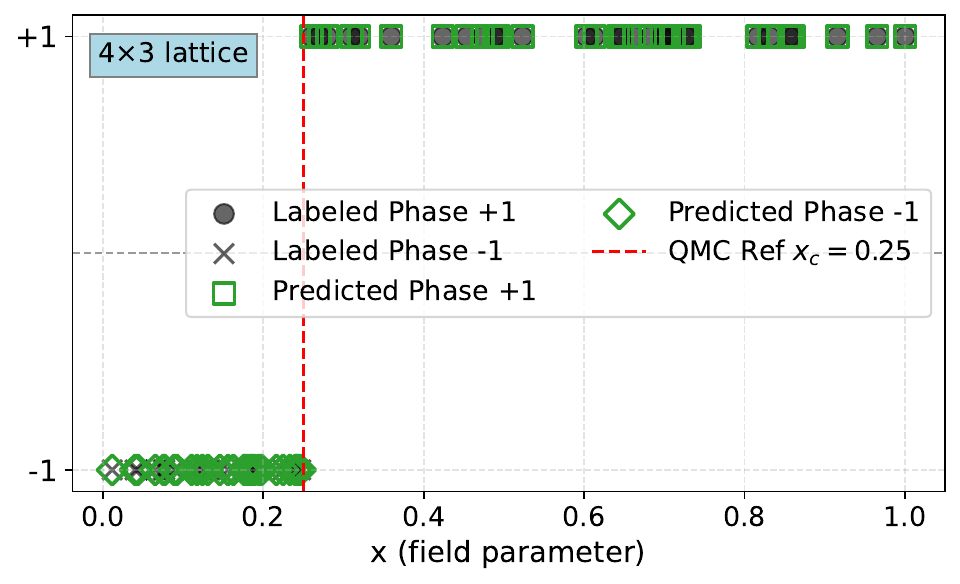}
    \end{subfigure}

    \caption{
    QCNN evaluation with randomly split datasets.
    Each panel corresponds to a different lattice size: \(2{\times}2\), \(2{\times}3\), \(3{\times}3\), and \(4{\times}3\).
    Predicted phase labels closely follow the true labels for all lattice sizes, demonstrating near-perfect classification accuracy across system sizes.
    }
    \label{fig:random_split}
\end{figure*}

\subsubsection{Unsupervised Learning: Quantum \texorpdfstring{$k$}{k}-means}
\label{sec:qkmeans}
In parallel, we explore an unsupervised approach based on quantum \(k\)-means clustering that relies on state overlaps~\cite{lloyd2013quantum,kerenidis2019q}. Given a set of ground states \(\{\ket{\psi_i}\}\), pairwise similarities are quantified by the state fidelity,
\begin{equation}
    F(\psi_i,\psi_j) = \big|\langle \psi_i | \psi_j \rangle \big|^2.
\end{equation}
From this, we define the Hilbert-Schmidt distance~\cite{Dodonov01032000} between pure states as 
\begin{equation}
    d(\psi_i,\psi_j) = \sqrt{2\big(1 - F(\psi_i,\psi_j)\big)} ,
\end{equation}
which provides a distance metric on the Hilbert space.
Clustering is performed with \(k=2\) by minimizing the total intra-cluster distance,
\begin{equation}
    \mathcal{L} = \sum_{c=1}^{k} \sum_{i \in C_c} d^2(\psi_i,\psi_{m_c}),
    \label{eq:dist}
\end{equation}
where \(\psi_{m_c}\) denotes the medoid state of cluster \(C_c\). Cluster labels are oriented such that small-\(x\) states correspond to the topological phase, ensuring consistent interpretation across system sizes. To identify the phase boundary, we analyze the cluster assignments as a function of the tuning parameter \(x\), similar to the QCNN approach. 
The estimated cluster flip interval point is then
\begin{equation}
    \hat{x}_c = \tfrac{1}{2}\big(x_{j_{\min}} + x_{j_{\max}}\big),
\end{equation}
with an associated uncertainty calculated identically to that used for the QCNN. Unlike the QCNN, quantum $k$-means requires no labeled data (and hence requiring no prior knowledge of the transition point even approximately) and partitions the quantum states directly into phases based solely on their pairwise overlaps, providing an unsupervised estimate of the transition point.

\begin{figure*}[t]
    \centering
    \includegraphics[width=0.98\linewidth]{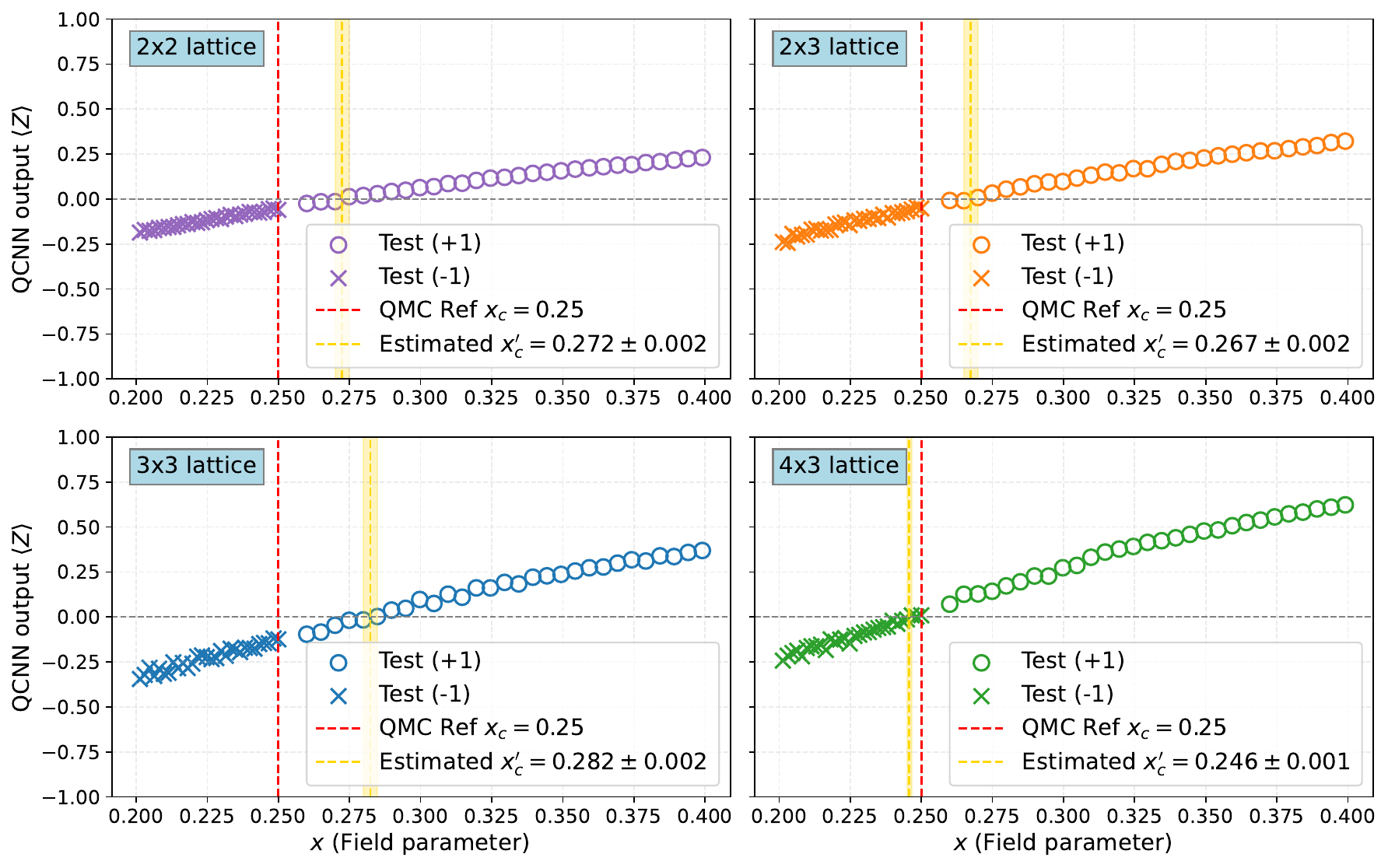}
    \caption{
    QCNN phase-boundary evaluation across lattice sizes.
    Raw QCNN outputs $\langle Z \rangle$ are shown as functions of the field parameter~$x$ for lattices 
    \(2{\times}2\), \(2{\times}3\), \(3{\times}3\), and \(4{\times}3\).
    The yellow-shaded regions indicate the phase flip-interval estimates of the transition point 
    \(x_c'\!\pm\!\Delta x\) extracted from the QCNN decision boundary, while the red dashed lines mark the 
    QMC benchmark \(x_c \approx 0.25\). As the system size increases, the QCNN-estimated phase boundary becomes close to the infinite-volume QMC critical point. %demonstrating scalability and physical consistency of the quantum convolutional architecture.
    }
    \label{fig:qcnn_physics_aware_raw_test}
\end{figure*}

\section{Experimental Results}
\label{sec:results}

\subsection{Toric-Code Dataset: Ground States from VQE}
\label{sec:results-vqe}
Using the procedure described in Sec.~\ref{sec:toric-code-ground}, we generate datasets of size 300 for lattices of size \(2{\times}2\), \(2{\times}3\), \(3{\times}3\), and \(4{\times}3\), corresponding to \(N=4\), 7, 12, and 17 qubits, respectively. Sampling is divided evenly between the two phases. Half of the states are prepared at equidistant field values \(x \in [0, x_c]\), covering the topologically ordered regime, and the other half at \(x \in [x_c + 10^{-2}, 1]\), covering the ferromagnetically ordered regime. This yields balanced datasets with uniform coverage of the parameter range.

To validate the accuracy of the generated states, we compare the VQE-optimized ground states against exact diagonalization (ED) results~\cite{Quspin2017}. The comparison includes the ground-state energy per qubit as
\begin{equation}
    E_N = \tfrac{1}{N}\langle \psi(\boldsymbol{\theta})|H_{\mathrm{TCM}}(x)|\psi(\boldsymbol{\theta})\rangle
\end{equation}
and the average magnetization per qubit as
\begin{equation}
    {\langle m_z \rangle}_N = \tfrac{1}{N}\sum_{i=1}^N \langle \sigma_i^z \rangle.
\end{equation}
Appendix~\ref{appx:vqe} shows that the per-qubit ground-state energies and average magnetizations obtained from VQE and exact diagonalization agree almost perfectly. 
Following Sec.~\ref{sec:toric-code-ground}, each ground state is assigned a phase label of \((-1)\) for the topologically ordered phase and \((+1)\) for the ferromagnetically ordered phase.

\subsection{Phase Learning via QCNN}
\label{sec:results-qcnn}

To learn phase classification, we first train the QCNN model described in Sec.~\ref{sec:qcnn} using a random 80\%–20\% train–test split across the full range of the field parameter \(x\). 
For each lattice size, 80\% of the toric-code states are used for training while the remaining 20\% are reserved for testing. 
The QCNN is optimized using the Adam optimizer with an initial learning rate of 0.01. 
Training is performed for 100 epochs with a batch size of 24 and includes an \(L_2\) regularization penalty with strength \(\lambda = 10^{-4}\) to suppress overfitting. 
More details on the training procedure and the learning curves are shown in Appendix~\ref{appx:training}.
After training, we evaluate the model on the testing states to compute the classification accuracy.

% To learn phase classification, we first train the QCNN model described in Sec.~\ref{sec:qcnn} using a random 80\%–20\% train–test split across the full range of the field parameter \(x\). For each lattice size, 80\% of the toric-code states are used for training while the remaining 20\% are reserved for testing. The QCNN is optimized using the Adam optimizer with an initial learning rate of 0.01. Training is performed for 100 epochs with a batch size of 24, including an \(L_2\) regularization penalty with strength \(\lambda = 10^{-4}\) to suppress overfitting. More detail on training 
% % \tb{Should we write out the cost function explicitly in Appendix~\ref{appx:training} and refer to it here? Physicists may not understand cross-entropy or \(L_2\).} 
% % The cost function is defined as the binary cross-entropy between predicted probabilities and labels, augmented by the \(L_2\) term. 
% More detail on the cost function and the learning curves are shown in Appendix~\ref{appx:training}. After training, we evaluate the model on the testing states to compute classification accuracy. 

\begin{figure*}[t]
    \centering
    \includegraphics[width=0.98\linewidth]{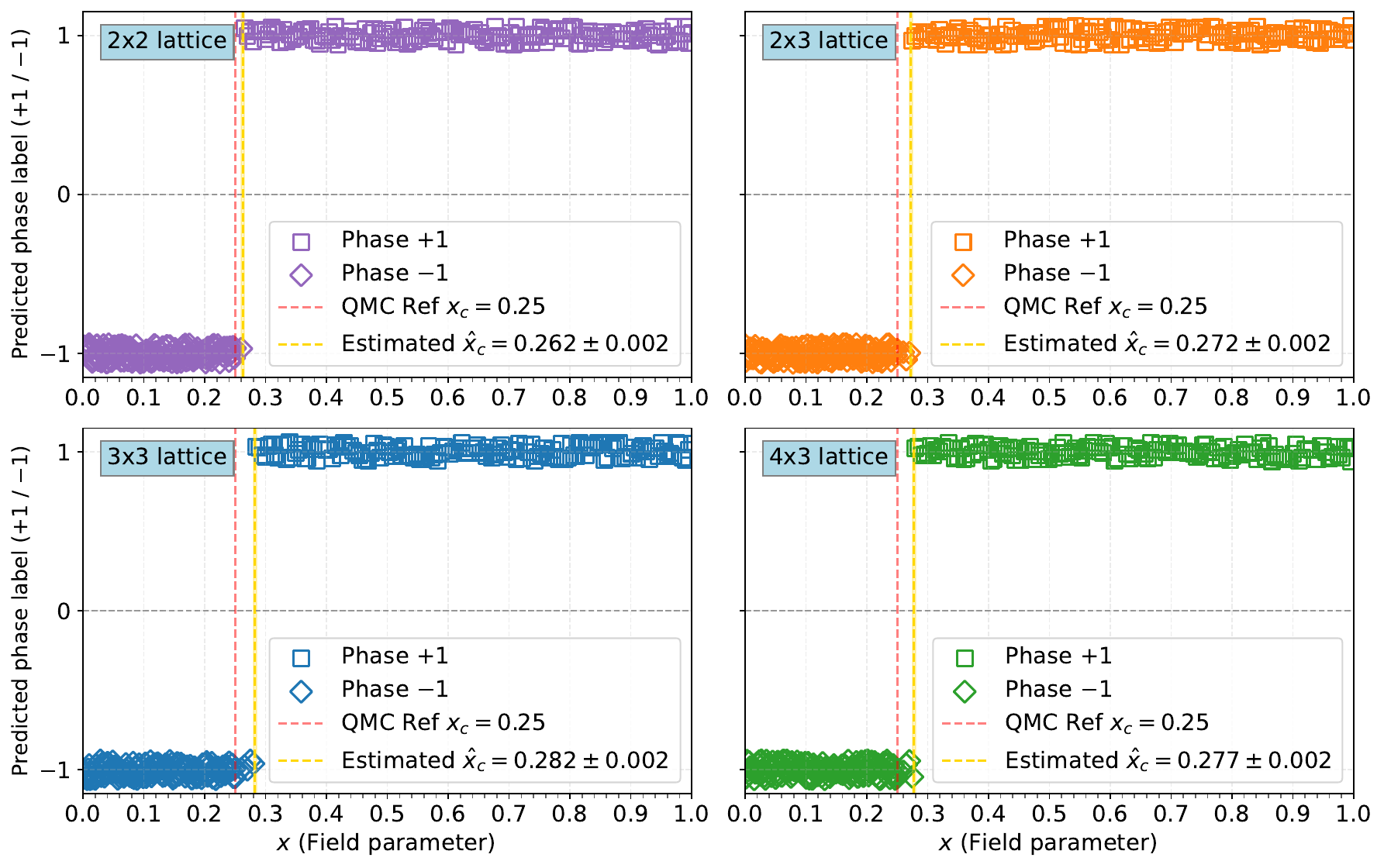}
    \caption{
    Quantum \(k\)-means phase classification of toric-code ground states.
    Each panel corresponds to a different lattice size: \(2{\times}2\), \(2{\times}3\), \(3{\times}3\), and \(4{\times}3\).
    The shaded yellow regions denote the phase transition intervals \(\hat{x}_c' \pm \Delta x\), while the red dashed line marks the QMC reference \(x_c \approx 0.25\).
    The cluster assignments (\(+1\) and \(-1\)) (jittered for visibility) exhibit clear phase separation and finite-size effects similar to the QCNN results~shown in Fig.~\ref{fig:qcnn_physics_aware_raw_test}. %indicating robust unsupervised detection of the topological transition.
    }
    \label{fig:qkmeans_phase_prediction}
\end{figure*}

As shown in Fig.~\ref{fig:random_split}, the predicted phase labels closely follow the true labels across all lattice sizes. The model achieves near-perfect classification accuracy: 100\% for the \(2{\times}2\), \(2{\times}3\), and \(4{\times}3\) lattices, and 96\% for the \(3{\times}3\) lattice. Each panel of Fig.~\ref{fig:random_split} corresponds to a specific lattice size, allowing direct comparison across system sizes and demonstrating excellent agreement between QCNN-predicted and labeled phases.

To investigate the phase-transition region, we adopt a physics-aware data split based on the QMC-estimated infinite-volume critical point \(x_c \approx 0.25\)~\cite{trebst2007,Wu2012phase}. In this setting, ground states away from the transition (\(x < 0.2\) or \(x > 0.4\)) are used for training, while those within the boundary window (\(0.2 \le x \le 0.4\)) are reserved for testing. This choice of boundary window ensures approximately balanced representations of both phases in the training and testing datasets. The training setup is same as random-split training discussed earlier. For each lattice size, a QCNN is trained on the off-critical data and evaluated on the test region. From the predicted phases in the boundary region, we also compute phase flip interval estimators $x_c'$ as discussed in Section~\ref{sec:qcnn}. 

The raw QCNN outputs as a function of \(x\) in the region (\(0.2 \le x \le 0.4\)) are shown in Fig.~\ref{fig:qcnn_physics_aware_raw_test}, where each subplot corresponds to one of the four lattice sizes (\(2{\times}2\), \(2{\times}3\), \(3{\times}3\), and \(4{\times}3\)). The vertical yellow dashed lines indicate the phase flip interval points $x_c'$, along with their range of variance, extracted from the QCNN decision boundary, while the red dashed line mark the infinite-volume QMC critical point \(x_c \approx 0.25\) for reference. 
%\tb{Delete this since the data doesn't allow such an unambiguous assertion. \st{We find that the QCNN phase flip interval estimates \hbox{\(x_c'\)} converge toward the QMC benchmark as the lattice size increases.}} 
For the \(2{\times}2\) lattice, the estimate is \(x_c' = 0.272(2)\), which decreases to \(x_c' = 0.267(2)\) for \(2{\times}3\). The \(3{\times}3\) system yields \(x_c' = 0.282(2)\), while the \(4{\times}3\) lattice produces \(x_c' = 0.246(1)\), in close agreement with the infinite-volume QMC value. Further study is needed to understand the nonmonotonic finite-volume effects implied by these observations. 
%\st{These results establish that the QCNN not only achieves accurate phase classification but also provides reliable finite-size estimates of the phase-transition point.}}

\subsection{Phase Detection via Quantum \texorpdfstring{$k$}{k}-Means}
\label{sec:results-qkmeans}
We next apply the unsupervised quantum \(k\)-means clustering procedure described in Sec.~\ref{sec:qkmeans} to the toric-code loop-gas datasets for each lattice size. Using the distance measure based on state overlaps as described in Eq.~\eqref{eq:dist}, the algorithm partitions the data into two clusters that are observed to correspond to the topological and ferromagnetic phases, without requiring any prior label information. We compute $\hat{x}_c$ as cluster flip interval point as discussed in Sec.~\ref{sec:qkmeans}. Fig.~\ref{fig:qkmeans_phase_prediction} shows the resulting phase clusters for all lattice sizes, together with the QMC reference point \(x_c\) and the cluster-flip estimates \(\hat{x}_c\).

The estimated cluster-flip points  \(\hat{x}_c\) are 0.262(2), 0.272(2), 0.282(2), and 0.277(2) for the \(2{\times}2\), \(2{\times}3\), \(3{\times}3\), and \(4{\times}3\) lattices, respectively, are similar to the QCNN phase flip estimates $x_c'$. This demonstrates that the phase structure can be extracted directly from quantum data in an unsupervised manner, perhaps with slight overestimation. Although in this small study the QCNN aligns more closely with the QMC benchmark, especially for the \(4{\times}3\) lattice, the quantum \(k\)-means approach remains a strong alternative given that no labels or prior knowledge are required.

\begin{figure*}[t!]
    \centering
    \begin{subfigure}{0.495\linewidth}
        \centering
        \includegraphics[width=\linewidth]{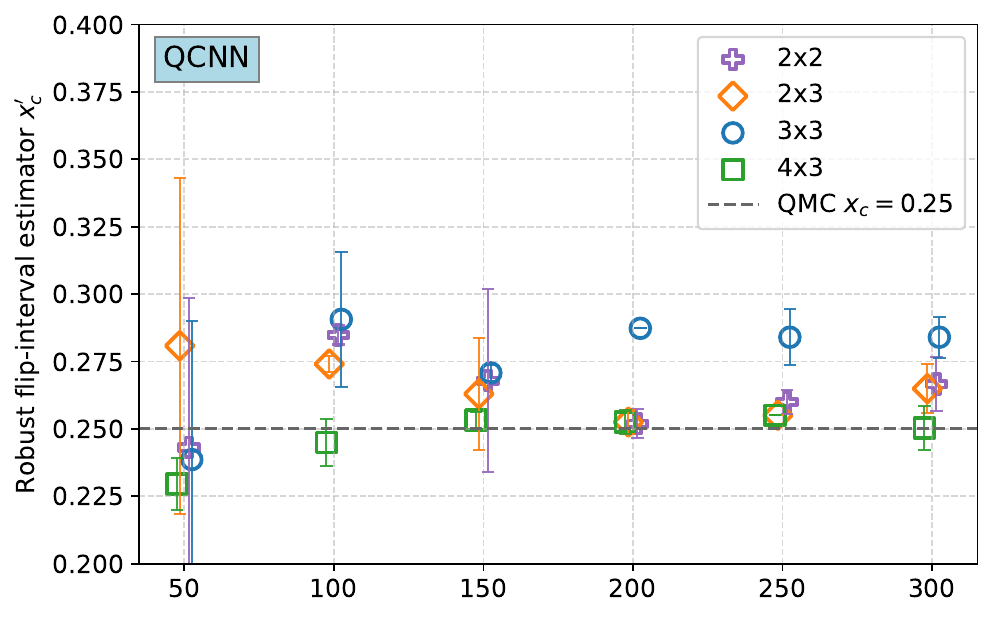}
    \end{subfigure}
    \hfill
    \begin{subfigure}{0.495\linewidth}
        \centering
        \includegraphics[width=\linewidth]{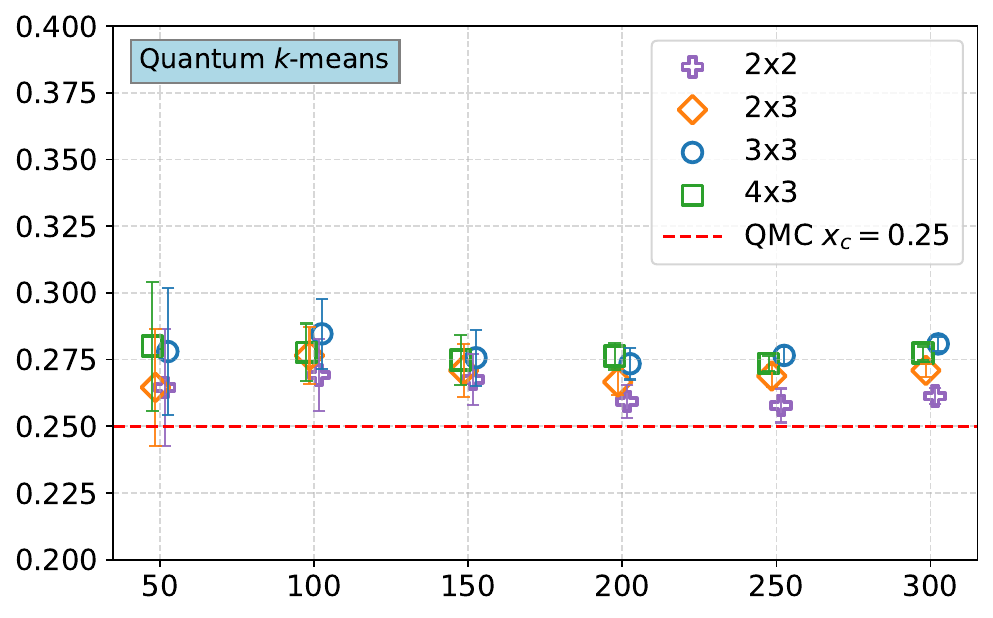}
    \end{subfigure}
    \begin{subfigure}{0.495\linewidth}
        \centering
        \includegraphics[width=\linewidth]{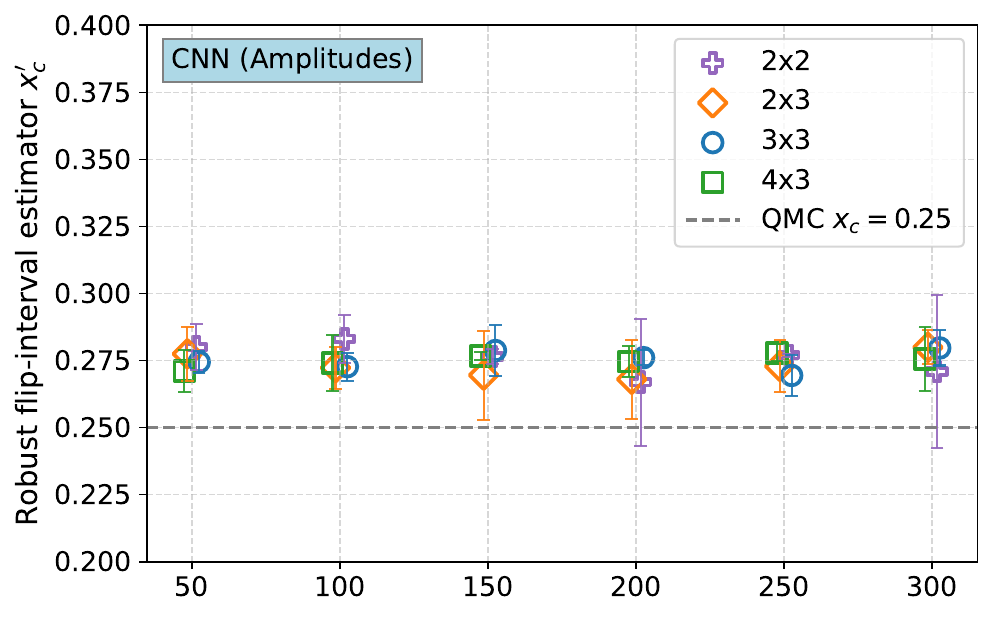}
    \end{subfigure}
    \hfill
    \begin{subfigure}{0.495\linewidth}
        \centering
        \includegraphics[width=\linewidth]{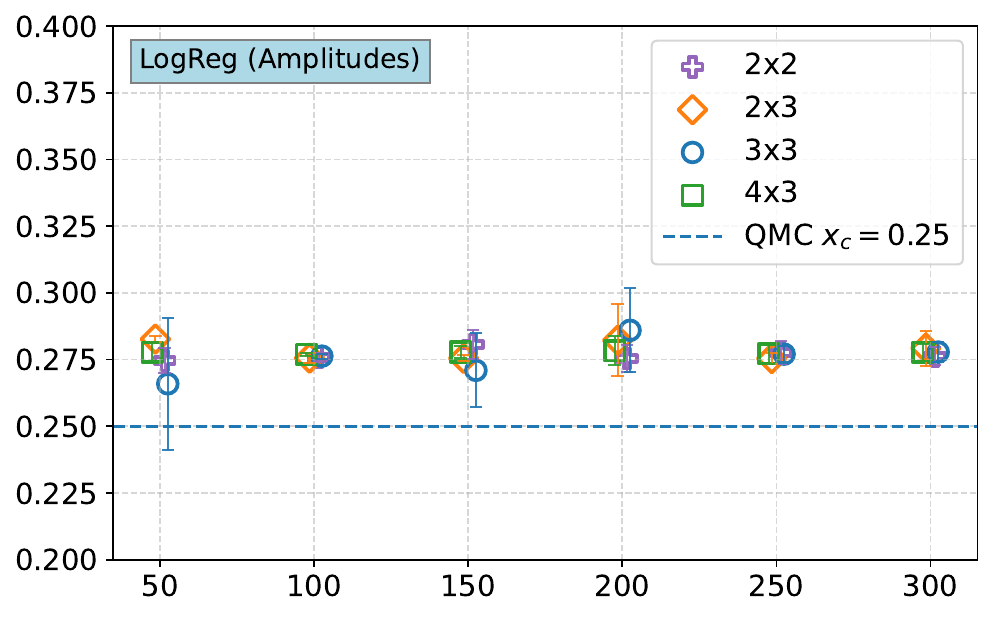}
    \end{subfigure}
    \begin{subfigure}{0.495\linewidth}
        \centering
        \includegraphics[width=\linewidth]{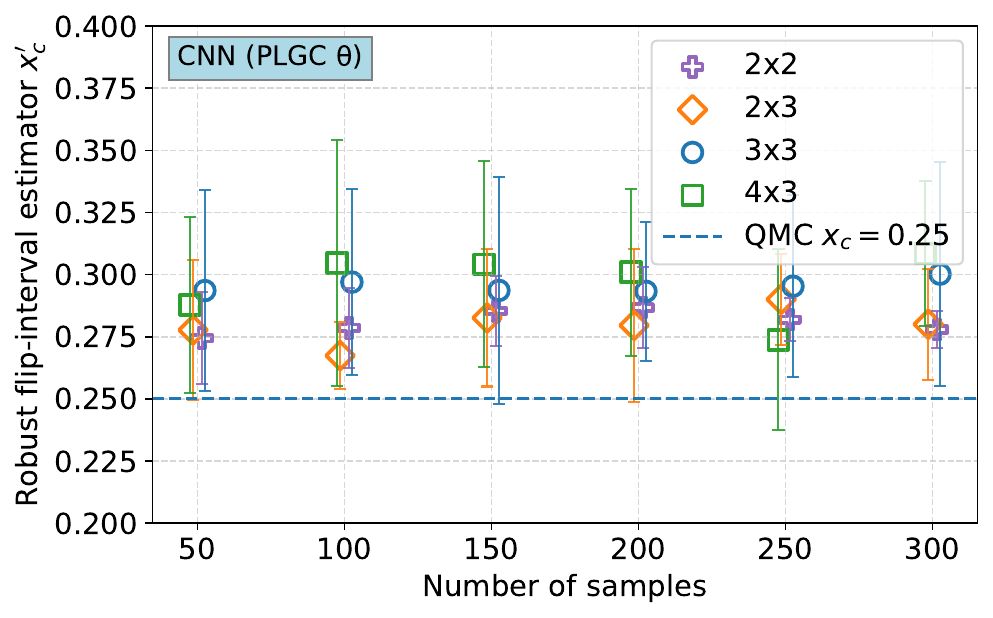}
    \end{subfigure}
    \hfill
    \begin{subfigure}{0.495\linewidth}
        \centering
        \includegraphics[width=\linewidth]{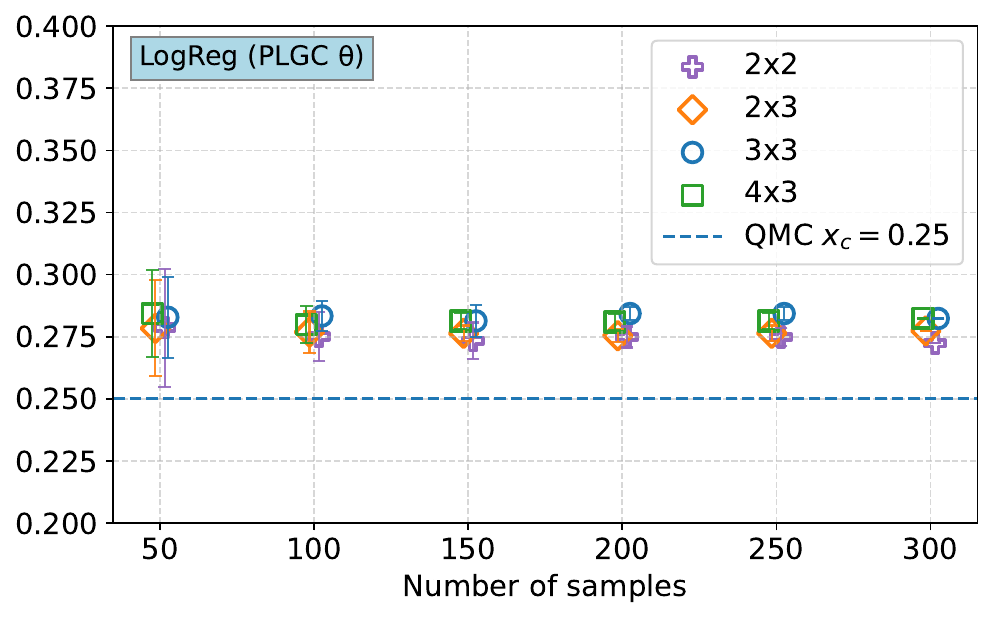}
    \end{subfigure}
    \caption{
        % \tb{Is it possible to stagger the points slightly in the horizontal direction to make the comparison more clearly visible?} Phase flip-interval estimator \(x_c'\) for phase-transition detection across quantum and classical learning models. 
        \textit{Top:} (\textit{left}) QCNN \(x_c'\); (\textit{right}) quantum \(k\)-means \(\hat{x}_c\). 
        \textit{Middle:} CNN (\textit{left}) and logistic regression (\textit{right}) trained on quantum-state amplitudes \(|\psi|^2\). 
        \textit{Bottom:} CNN (\textit{left}) and logistic regression (\textit{right}) trained on PLGC parameters \(\boldsymbol{\theta}\) (parameter-based models). 
        Each subplot shows the estimated phase interval point as a function of training-sample size for lattices \(2{\times}2\), \(2{\times}3\), \(3{\times}3\), and \(4{\times}3\). For better visibility, a small horizontal offset is applied to the data points along the \(x\)-axis.
        Classical models exhibit systematic shifts from the QMC benchmark (\(x_c \approx 0.25\)) of a size comparable to the shift seen in our QCNN experiment. The QCNN converges toward \(x_c\) for larger lattices even with \(200\) training samples while the quantum \(k\)-means model closely tracks the phase boundary without supervision.
    }
    \label{fig:classical_vs_quantum}
\end{figure*}

\subsection{Comparison with Classical Machine-Learning Models}
\label{sec:results-classical}

To benchmark the QDL approaches against classical counterparts, we train convolutional neural networks (CNNs) and logistic regression classifiers on the same toric-code loop-gas datasets. 
Both classical models take as input either the state amplitudes \(|\psi_x|^2\) in the computational basis or the corresponding VQE ansatz parameters \(\boldsymbol{\theta}\). 
The dimensionality of the amplitude vectors grows exponentially with the number of qubits, becoming computationally intractable for larger lattice sizes. Using the variational parameters \(\boldsymbol{\theta}\) as input is also not ideal, since these parameters are not directly accessible for an arbitrary ground state. Previous studies have explored classical shadows or sampled representations as alternative inputs, but such approaches require additional post-processing~\cite{huang2020predicting,huang2022learning,huang2022provably, struchalin2021experimental, zhang2021experimental}. Since the QCNN operates directly on quantum states, our comparison focuses exclusively on models trained with either state amplitudes or VQE ansatz parameters as input.

We compare the phase flip-interval estimations obtained from the QCNN, quantum \(k\)-means, CNN, and logistic regression (LR) models. 
Each model is trained on randomly selected subsets of 50, 100, 200, and 300 samples drawn from the off-critical regions of the magnetic-field parameter \(x\), while reserving 300 samples within the boundary window (\(0.2 \leq x \leq 0.4\)) for testing. 
This setup ensures that all models are trained on balanced phases within sampled states and are evaluated on the total unseen states near the phase transition. 
Each training procedure is repeated ten times, and the resulting phase flip-interval estimates are reported as the mean and standard deviation across repetitions.

Fig.~\ref{fig:classical_vs_quantum} presents a comparison of the average phase flip-interval estimations between the quantum and classical models. The top panels show the QCNN estimates \(x_c'\) (\textit{left}) and the quantum \(k\)-means estimates \(\hat{x}_c\) (\textit{right}). 
The middle panels display the estimated \(x_c'\) for the CNN and logistic regression models trained on quantum-state amplitudes \(|\psi|^2\), while the bottom panels correspond to CNN and logistic regression models trained on the parametrized loop-gas circuit (PLGC) parameters \(\boldsymbol{\theta}\). Each subplot presents the estimated transition point as a function of training-sample size for lattice geometries up to \(4{\times}3\).

\begin{figure*}[t]
    \centering
    \includegraphics[width=0.8\linewidth]{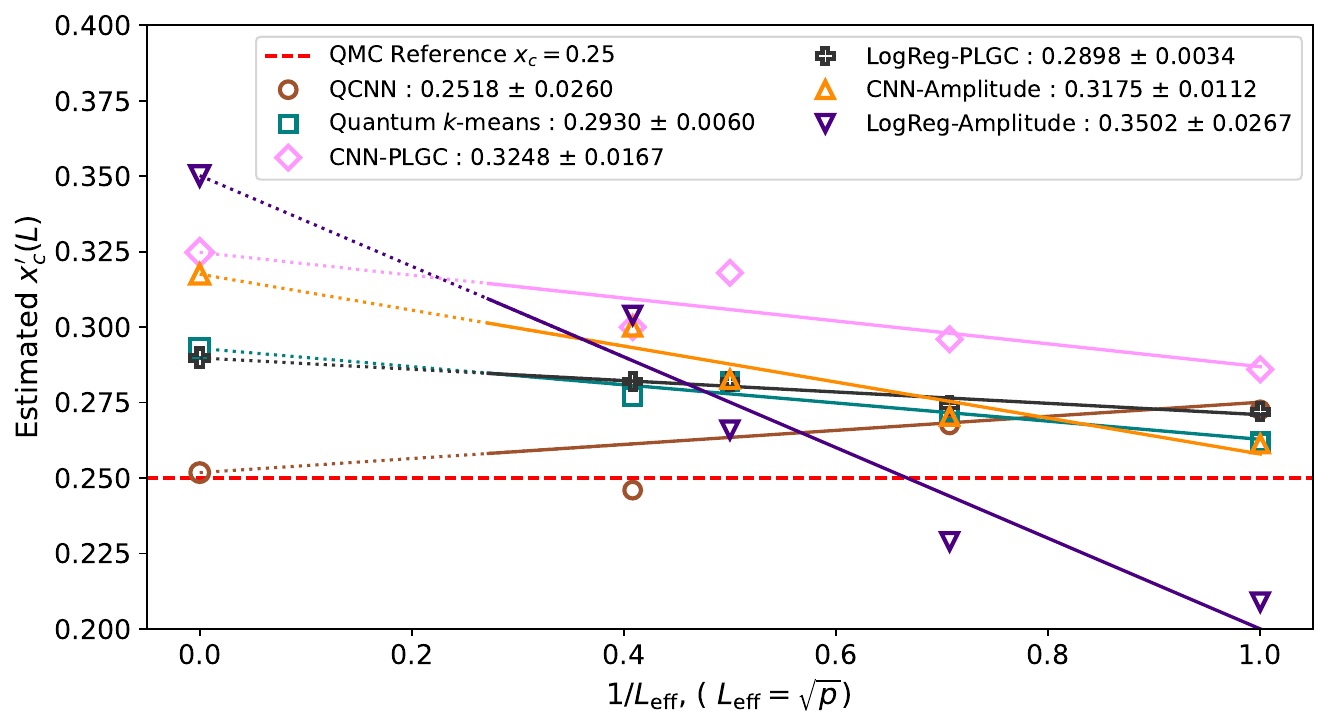}
    \caption{
    Finite-size scaling of estimated critical points \(x_c'(L)\) versus \(1/L_{\mathrm{eff}}\) for QCNN, Quantum \(k\)-means, CNN, and logistic regression.
    Solid lines denote fits to the scaling form \(x_c'(L) = x_c(\infty) + a/L_{\mathrm{eff}}\), and markers indicate results for each lattice size.
    The red dashed line shows the QMC reference \(x_c \approx 0.25\).
    }
    \label{fig:finite_size_scaling}
\end{figure*}

For the QCNN, the phase estimates converge toward \(x_c \approx 0.25\) for larger lattice sizes with relatively few samples compared to the classical models. The \(3{\times}3\) system exhibits a slight upward shift in \(x_c'\), while the \(4{\times}3\) lattice converges near the QMC benchmark with as few as 100 samples. For the quantum \(k\)-means algorithm, the cluster-flip estimates \(\hat{x}_c\) remain slightly above the QMC value (0.26–0.29) yet stable with increasing sample size. In contrast, both CNN and logistic regression models show less consistent and more erratic learning behavior for both amplitude-based and VQE-parameter-based inputs. Their phase flip estimates also systematically overestimate the transition point relative to the QMC reference. We further investigate both classical models and find that their predictions fluctuate within the testing region \(0.2 \leq x \leq 0.4\). Appendix~\ref{appx:training} shows the phase-label prediction plots for both classical models considered in this comparison in Fig.~\ref{fig:classical_models_phase_prediction_4x3}. We also find that classical models trained on PLGC inputs exhibit smaller fluctuations in phase prediction than those trained on amplitudes. This indicates that the PLGC representation encodes more robust phase-related information than the amplitude representation. Overall, we conclude that the QDL models demonstrate the ability to learn phase boundaries directly from quantum states without requiring order parameters or prior labeling as well as or better than the classical approaches.

% CNNs trained on PLGC variational parameters \(\boldsymbol{\theta}\) show closer convergence toward the QMC critical value for the \(4{\times}3\) lattice with 300 samples. Overall, models trained on PLGC parameters \(\boldsymbol{\theta}\) demonstrate improved learning stability and convergence with increasing dataset size. This improvement suggests that the PLGC representation encodes richer, physically structured information about the underlying phase transition, enabling classical models to approximate the quantum decision boundary more accurately. However, such parameters are accessible only when the ground state is generated through a parametrized circuit like the PLGC. For arbitrary toric-code ground states, these parameters are not directly measurable or reconstructable from the quantum state.

%%%%Will work from here next
%% CNN and LR Phase label plot

\subsection{Finite-Size Scaling}
\label{sec:finite-size-scaling}

To consolidate the results across different lattice sizes, we perform a finite-size scaling analysis of the estimated phase-transition points \(x_c'(L)\). 
These estimates are obtained from QDL techniques, including the QCNN and Quantum \(k\)-means, as well as from classical models (CNN and logistic regression) trained on either VQE ansatz parameters or amplitude representations.

For each method, the critical-point estimates are fitted to the empirical scaling relation
\begin{equation}
    x_c'(L) = x_c(\infty) + \frac{a}{L_{\mathrm{eff}}},
\end{equation}
where \(L_{\mathrm{eff}} = \sqrt{p}\) with \(p\) the number of plaquettes, and \(x_c(\infty)\) denotes the extrapolated critical point in the thermodynamic limit. The lattice sizes used in this study (\(2\times2\), \(2\times3\), \(3\times3\), and \(4\times3\)) contain \(p = 1, 2, 4,\) and \(6\) plaquettes, respectively. Fig.~\ref{fig:finite_size_scaling} summarizes the resulting finite-size behavior for all models.

Even though the \(3\times3\) lattice seems to have an anomalously large finite size effect, the QCNN yields an extrapolated critical point of \(x_c(\infty) = 0.2518 \pm 0.0260\), which lies within one standard deviation of the QMC benchmark \(x_c \approx 0.25\). 
Its one-sigma confidence interval, \([0.226,\,0.278]\), comfortably includes the QMC value, and the small fitting uncertainty indicates stable convergence across lattice sizes. 
The Quantum \(k\)-means method gives \(x_c(\infty) = 0.2930 \pm 0.0060\), slightly overestimating the transition. 
Among the classical approaches, the CNN trained on PLGC inputs extrapolates to \(x_c(\infty) = 0.3248 \pm 0.0167\), compared with \(x_c(\infty) = 0.3175 \pm 0.0112\) for the amplitude-based CNN. 
Logistic regression follows a similar trend, yielding \(x_c(\infty) = 0.2898 \pm 0.0034\) for PLGC inputs and \(x_c(\infty) = 0.3502 \pm 0.0267\) for amplitude inputs. 
Both classical models exhibit significant deviations from the QMC benchmark \(x_c \approx 0.25\).
Overall, in our experiments, the QCNN provides the most accurate estimation, closely reproducing the thermodynamic transition point within its one-sigma confidence range.

\section{Conclusion}
\label{sec:conclusion}

Experimental results demonstrate that QDL provides an effective framework for probing the topological-to-ferromagnetic phase transition in the \(2{+}1\)-dimensional toric-code loop-gas model. Quantum convolutional neural networks (QCNNs) achieve near-perfect phase classification and yield phase flip estimates that converge toward the QMC benchmark, with finite-size scaling giving \(x_c(\infty) = 0.2518 \pm 0.0260\) in excellent agreement with the previously known result \(x_c \approx 0.25\). Remarkably, QCNNs exhibit robust scaling behavior even when trained on data from only four modest lattice sizes, underscoring their ability to extract critical behavior from limited quantum resources. The Quantum \(k\)-means clustering method similarly identifies the phase boundary without requiring labeled data, albeit with a small upward shift, while classical CNN and logistic-regression models consistently overestimate the transition point.

The main limitation of the present study lies in the modest lattice sizes (up to \(4{\times}3\)) accessible by classical simulation, where finite-size fluctuations remain non-negligible. Future extensions will aim to scale QDL approaches to larger lattices and more complex models with intrinsic topological order, supported by improved state-preparation techniques and hardware-efficient implementations on near-term quantum devices.

Building on these results, our QDL study of the toric-code transition suggests a promising route toward learning phase diagrams of lattice gauge theories relevant to quantum chromodynamics (QCD). The QCD phase diagram features confinement–deconfinement and chiral transitions, potential critical endpoints, and intricate finite-density and finite-temperature behavior that remain difficult to access with classical methods~\cite{Aarts_2023}. As an intermediate, controllable step, a \(\mathbb{Z}_2\) lattice gauge theory with dynamical fermions~\cite{Homeier_2023} provides a natural testbed for QDL-based classifiers. Its phase structure—encompassing confined “hadronic” regimes, deconfined or quark-liquid–like sectors, and superfluid or Mott-insulating phases depending on the coupling and chemical potential—shares qualitative features with QCD and is well suited for exploration using supervised QCNNs and unsupervised, overlap-based clustering methods aimed at detecting first-order transitions, crossovers, and critical points.

Extending beyond the ground-state transitions explored here, the study of thermal transitions in such gauge–matter systems will require the preparation of mixed quantum states. Recently developed thermal pure quantum (TPQ) methods for gauge theories~\cite{Davoudi_2023} enable efficient construction of finite-\(T\) and finite-\(\mu\) states on quantum hardware, providing a natural foundation for applying QDL to finite-temperature and finite-density physics. Together, these directions outline a broader program for employing QDL architectures to map the phase structure of strongly correlated gauge theories and to bridge quantum simulation, quantum information, and high-energy physics.

\section{Acknowledgements}
The research presented in this article was supported by the NNSA’s Advanced Simulation and Computing Beyond Moore’s Law program at Los Alamos National Laboratory and the Laboratory Directed Research and Development program of Los Alamos National Laboratory under project number 20260043DR. 
This work has been assigned LANL technical report number LA-UR-25-30864.

\begin{figure}
    \centering
    \includegraphics[width=0.95\linewidth]{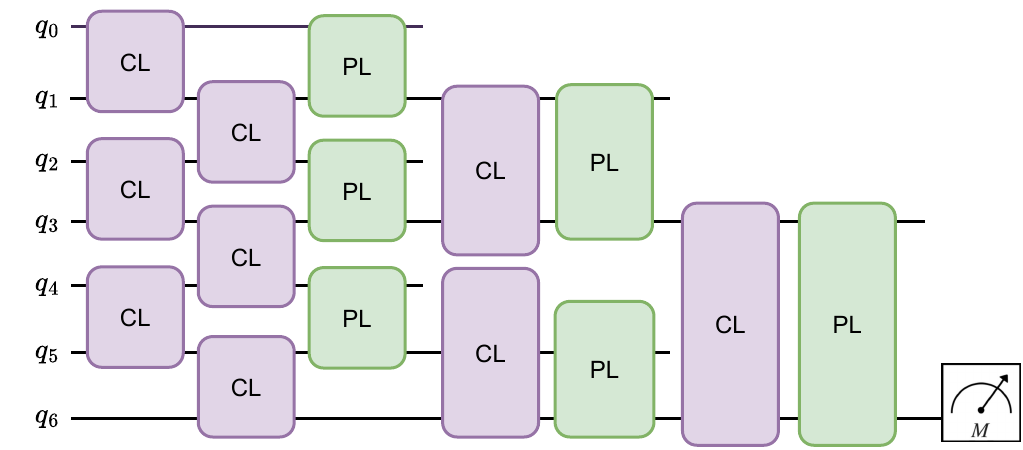}
    \caption{
    Schematic QCNN network topology for the \(2\times3\) lattice.
    Alternating convolution (CL) and pooling (PL) layers progressively reduce 
    the number of active qubits, followed by a final measurement on the last qubit.
    }
    \vspace*{-2.25\baselineskip}
    \label{fig:qcnn-schematic}
\end{figure}

\appendix
\section{VQE Training}
\label{appx:vqe}

We employed the PLGC ansatz within the VQE framework to generate toric-code ground states for $x\in[0,1]$. Optimization is  performed using the simultaneous perturbation stochastic approximation (SPSA) algorithm~\cite{sapsa1992} with a learning rate of $0.01$. Each VQE run was initialized with random angles \(\theta_p \in [0,2\pi)\), and multiple independent trials (ten per \(x\) value) were performed to mitigate convergence to local minima. Each trial proceeded for $500$ optimization steps, and the parameters yielding the lowest final energy were retained.  We computed the ground-state energy per qubit and average magnetization per qubit, \(\langle m_z \rangle\), defined in Eq.~\eqref{eq:magnetization_per_qubit}, for all lattice sizes.
%which serves as a complementary diagnostic of the phase transition. \tb{In what sense are they diagnostic of phase transtion? There is nothing remarkable that happens in Fig.~\ref{fig:vqe_vs_ed_combined} near \(x\approx0.25\)?}
\begin{figure*}[t]
    \centering
    \begin{subfigure}{0.45\linewidth}
        \centering
        \includegraphics[width=\linewidth]{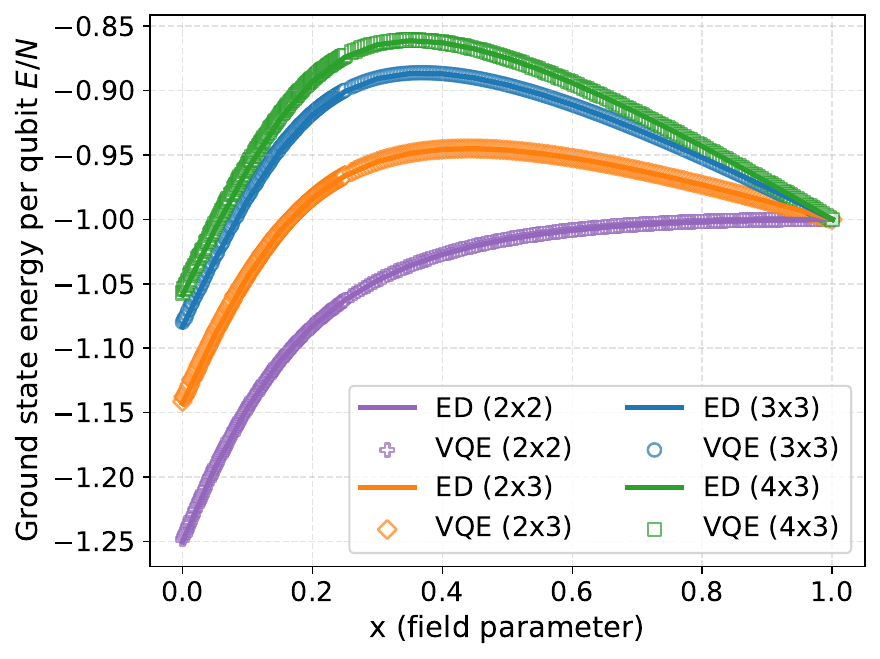}
        \label{fig:vqe_vs_ed}
    \end{subfigure}\hfill
    \begin{subfigure}{0.45\linewidth}
        \centering
        \includegraphics[width=\linewidth]{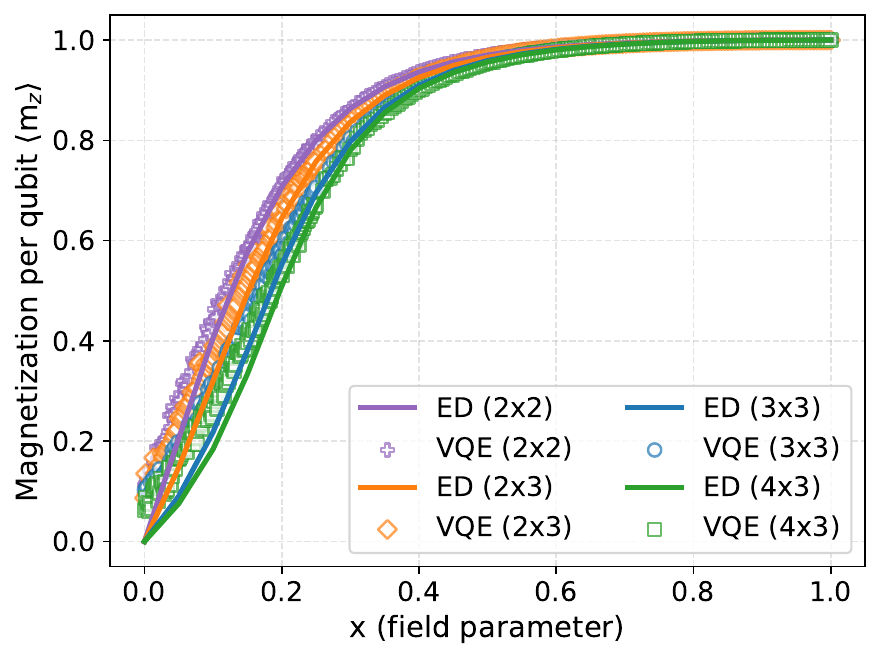}
        \label{fig:vqe_vs_ed_mz}
    \end{subfigure}
    \caption{Comparison between exact diagonalization (ED, solid lines) and variational quantum eigensolver (VQE) simulations using the parametrized loop-gas circuit (PLGC) for lattice sizes \(2{\times}2\), \(2{\times}3\), \(3{\times}3\), and \(4{\times}3\). (a) Ground-state energy per qubit \(E/N\) as a function of the tuning parameter \(x\). (b) Average magnetization per qubit \(\langle m_z \rangle = N^{-1}\sum_i\langle \sigma_i^z \rangle\). The close agreement demonstrates the accuracy of VQE-generated states in reproducing both energy and order parameters across system sizes.}
    \label{fig:vqe_vs_ed_combined}
\end{figure*}
Fig.~\ref{fig:vqe_vs_ed_combined} compares the ground states obtained from exact diagonalization (ED) with those generated by VQE across different lattice sizes. The left panel shows that the energy per qubit from VQE \textit{(symbols)} is nearly indistinguishable from the ED results \textit{(solid lines)} over the full field range, while the right panel shows similarly close agreement in the average magnetization. Both observables capture the transition from the low-field topological phase to the high-field ferromagnetic phase. These results suggest that the PLGC ansatz accurately reproduces the ground-state physics of the toric-code model under a magnetic field, even for larger lattices where ED becomes computationally expensive.

\begin{figure}
    \centering
    \scriptsize
    % ---------- (a) CL block ----------
    \begin{subfigure}{0.95\linewidth}
        \centering
        \begin{quantikz}[thin, row sep=0.12cm, column sep=0.15cm]
        \lstick{$q_0$} & \gate{R_z(\alpha_1)} & \gate{R_y(\alpha_2)} & \gate{R_z(\alpha_3)} & \ctrl{1} & \qw & \ctrl{1} & \qw \\
        \lstick{$q_1$} & \gate{R_z(\alpha_4)} & \gate{R_y(\alpha_5)} & \gate{R_z(\alpha_6)} & \targ{} & \gate{R_y(\alpha_7)} & \targ{} & \gate{R_y(\alpha_8)}
        \end{quantikz}
        \vspace{-3pt}
        \caption*{\textit{(a)} Convolution Block}
    \end{subfigure}

    \vspace{8pt}

    % ---------- (b) PL block ----------
    \begin{subfigure}{0.95\linewidth}
        \centering
        \begin{quantikz}[thin, row sep=0.12cm, column sep=0.16cm]
        \lstick{$q_0$} & \gate{R_y(\beta_1)} & \gate{R_z(\beta_{2})} & \ctrl{1} & \qw \\
        \lstick{$q_1$} & \gate{R_y(\beta_{3})} & \gate{R_z(\beta_{4})} & \targ{} & 
                          \gate{R_z(-\beta_{4})} & \gate{R_y(-\beta_{3})} & \qw
        \end{quantikz}
        \vspace{-3pt}
        \caption*{\textit{(b)} Pooling Block}
    \end{subfigure}

    \caption{
        QCNN building blocks.
        \textit{(a)} Convolution layer (CL) implementing an $SU(4)$-like two-qubit unit.
        \textit{(b)} Pooling layer (PL) transferring correlations to the control qubit and discarding redundant degrees of freedom.
    }
    \label{fig:qcnn-circuit-blocks}
    \vspace*{-3\baselineskip}
\end{figure}
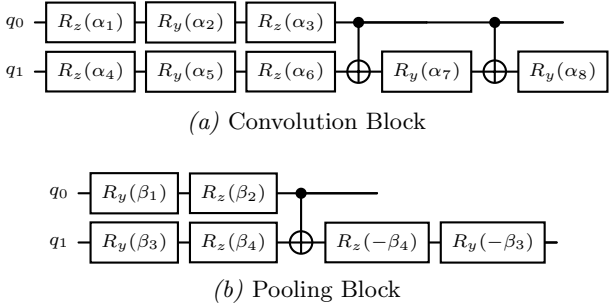

\section{QCNN Model Architecture}
\label{appx:qcnn}
The QCNN model used in this work consists of alternating convolution (CL) and pooling (PL) layers designed to hierarchically coarse-grain quantum information encoded in the toric-code ground states~\cite{nagano2023quantum}. 
Fig.~\ref{fig:qcnn-schematic} illustrates the QCNN architecture, which comprises seven qubits shown in Fig.~\ref{fig:toric_code_qdl} corresponding to the \(2\times3\) lattice. 
Each convolution layer applies a set of parameterized two-qubit \(SU(4)\) gates arranged in a brickwork pattern, followed by pooling operations that progressively reduce the number of active qubits from seven to four, then to two, and finally to one. 
The CL blocks perform local entangling transformations that extract multiqubit correlations and emulate the action of short-range renormalization operators, while the PL blocks implement nonunitary coarse-graining through measurement-inspired controlled rotations that transfer relevant information to the control qubits while discarding redundant degrees of freedom. 
Fig.~\ref{fig:qcnn-circuit-blocks} shows the internal circuits of the convolution block (top) and the pooling block (bottom), parameterized by the learnable parameters~$\{\alpha_i\}$ and~$\{\beta_i\}$ .

\section{Learning Phase Transition}
\label{appx:training}
\begin{figure*}[t!]
    \centering
    \includegraphics[width=0.95\linewidth]{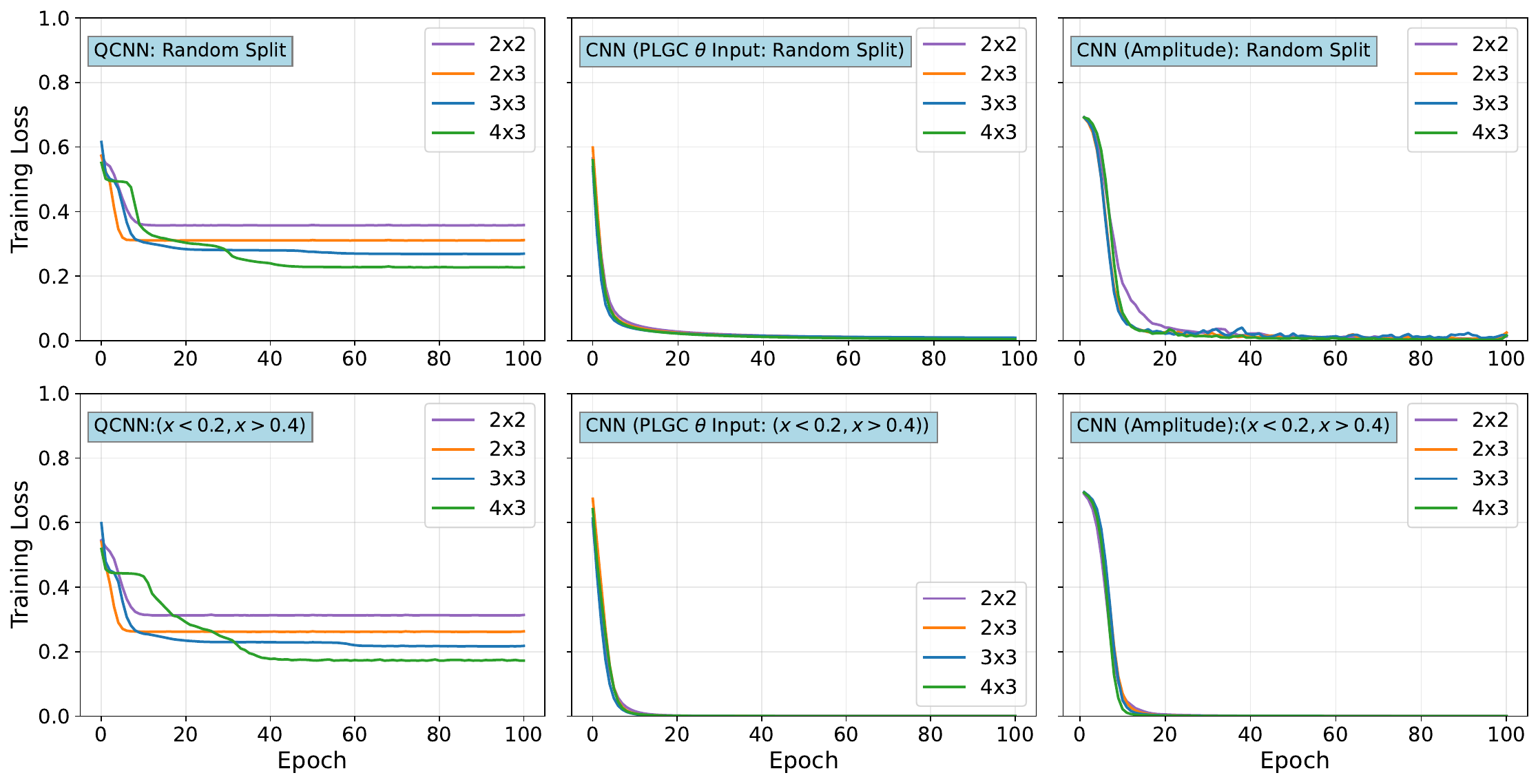}
    \caption{
    Comparison of training losses for QCNN and CNN models.
    \textit{Top:} QCNN training under random-split data (left) and a physics-aware split excluding the near-critical region (\(0.2 \le x \le 0.4\)) (right).
    \textit{Bottom:} CNN training using the same random (left) and physics-aware (right) splits.
    In both architectures, the physics-aware split yields consistently lower losses and smoother convergence, as data outside the critical window are more easily separable into distinct phases.
    The CNN converges more rapidly toward zero loss, reflecting its simpler classical optimization landscape compared with the QCNN.
    }
    \label{fig:qcnn_training_loss}
\end{figure*}

We train both QCNN and CNN models to learn phase classification and to estimate the phase-transition region. 
Each model is trained on a randomly selected 80\% subset of ground states from the dataset and evaluated on the remaining 20\%. 
Fig.~\ref{fig:qcnn_training_loss}~\textit{(top)} shows the training losses for both the QCNN~\textit{(left)} and the CNN models trained on either VQE ansatz parameters~\textit{(middle)} or state amplitudes~\textit{(right)}. 
The plots indicate that the CNN models converge faster and achieve lower training losses than the QCNN, reflecting the simpler optimization landscape of classical architectures. 

The QCNN is optimized using the Adam algorithm with an initial learning rate of 0.01, batch size of 24, and an \(L_2\) regularization strength \(\lambda = 10^{-4}\) to suppress overfitting. 
The cost function used for training is the binary cross-entropy between predicted probabilities and target phase labels, augmented by an \(L_2\) penalty on the QCNN parameters,
\begin{equation}
    \mathcal{L} = 
    -\frac{1}{N}\sum_i 
    \Big[y_i \log p_i + (1 - y_i) \log(1 - p_i)\Big]
    + \lambda \frac{\|\boldsymbol{\phi}\|_2^2}{N},
    \label{eq:qcnn_loss}
\end{equation}
where \(p_i = (1 + y_{\mathrm{out}}(\boldsymbol{\phi})) / 2\) is the predicted probability derived from the QCNN output 
\(y_{\mathrm{out}}(\boldsymbol{\phi}) = \langle Z_{\mathrm{out}} \rangle_i\) 
as defined in Eq.~\eqref{eq:qcnn_output}, and \(y_i \in \{0,1\}\) denotes the target phase label. 
The first term measures the misclassification error, while the \(L_2\) term regularizes the trainable parameters \(\boldsymbol{\phi}\) to improve generalization and prevent overfitting.

Training is performed for 100 epochs, during which the learning rate is adaptively decayed according to a step schedule to ensure stable convergence. 
Optimization is monitored through the mean loss per epoch, and convergence is reached when successive loss values vary by less than \(10^{-3}\). 

Training is performed for 100 epochs, during which the learning rate is adaptively decayed according to a step schedule to ensure stable convergence. 
Optimization is monitored through the mean loss per epoch, and convergence is reached when successive loss values vary by less than \(10^{-3}\). 
The training curves in Fig.~\ref{fig:qcnn_training_loss} illustrate this behavior.

To estimate the phase transition, each model is trained on ground states away from the critical region (\(x < 0.2\) or \(x > 0.4\)) and evaluated within the testing region (\(0.2 \leq x \leq 0.4\)). 
As shown in Fig.~\ref{fig:qcnn_training_loss}~\textit{(bottom)}, the physics-aware split yields lower losses and smoother convergence than the random split, since training data in the critical window are potentially corrupted by finite volume effects. For this split, the CNN models also demonstrate faster convergence than the QCNN.

\begin{figure*}[t!]
    \centering
    \includegraphics[width=0.95\linewidth]{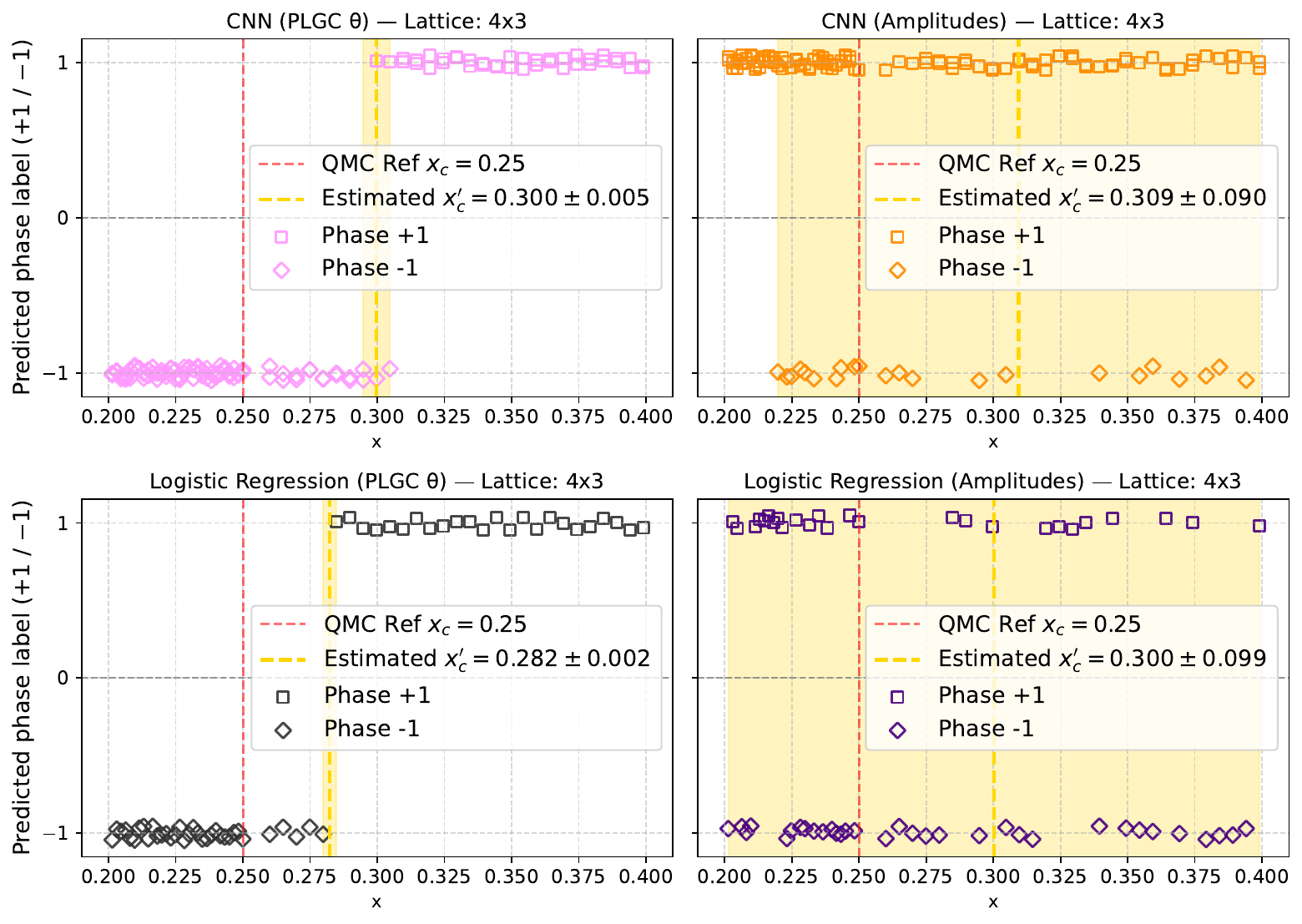}
    \caption{
        Phase-label predictions (+1 and -1) from classical machine-learning models (jittered for visibility)
        trained on toric-code quantum-state datasets for the \(4{\times}3\) lattice. 
        The top panel shows phase predictions from CNNs
        using input PLGC parameters \(\boldsymbol{\theta}\) \textit{(left)} and state amplitudes \(|\psi|\) \textit{(right)}. 
        The bottom panel shows phase predictions from logistic-regression models
        using the same inputs, PLGC parameters \(\boldsymbol{\theta}\) \textit{(left)} and amplitudes \(|\psi|\) \textit{(right)}.
        The red dashed line denotes the QMC reference
        (\(x_c \approx 0.25\)), while the yellow dashed line marks the flip interval estimator
        \(x_c'\) extracted from each model.
        Models trained on PLGC inputs yield more accurate phase boundaries with a smaller range of variance 
        than those trained on amplitudes, although both deviate noticeably 
        from the QMC benchmark.
    }
    \label{fig:classical_models_phase_prediction_4x3}
\end{figure*}

For both classical and quantum models, the phase transition point is determined from the model-predicted phase labels. 
Fig.~\ref{fig:classical_models_phase_prediction_4x3} shows phase-label predictions from classical machine-learning models: CNNs (\textit{top}) and logistic regression (\textit{bottom}) for the \(4{\times}3\) lattice. 
The results show that models trained on VQE ansatz parameters (\textit{left}) yield more accurate phase estimations than those trained on state amplitudes (\textit{right}). 
In contrast, both CNN and logistic-regression models trained on amplitudes exhibit fluctuating and less stable predictions within the testing region. 
Although the loss curves in Fig.~\ref{fig:qcnn_training_loss} indicate better convergence for the CNN models on training states, they still show poor generalization around the phase boundary.

\bibliography{main}% Produces the bibliography via BibTeX.
\end{document}